\documentclass[letterpaper,peerreview,draftcls,onecolumn,11pt]{IEEEtran}
\usepackage{amsmath, amssymb, epsfig, graphicx, psfrag, subfigure, itrans}

\title{Malleable Coding: Compressed Palimpsests}%
\author{Lav R. Varshney, \IEEEmembership{Graduate Student Member, IEEE,}
        Julius Kusuma, \IEEEmembership{Member, IEEE,} \\
        and Vivek K Goyal, \IEEEmembership{Senior Member, IEEE}%
\thanks{This work was supported in part by the NSF Graduate
	  Research Fellowship, Grant CCR-0325774, and Grant CCF-0729069.}
\thanks{L.~R.~Varshney is with the Department of Electrical Engineering and Computer Science, 
	  the Laboratory for Information and Decision Systems, and
	  the Research Laboratory of Electronics, Massachusetts Institute of Technology,
	  Cambridge, MA  02139 USA (e-mail: lrv@mit.edu).}%
\thanks{J.~Kusuma is with Schlumberger Technology Corporation, Sugar Land, TX 77478 USA (e-mail: kusuma@alum.mit.edu).}%
\thanks{V. K. Goyal is with the Department of Electrical Engineering and Computer Science and
	  the Research Laboratory of Electronics, Massachusetts Institute of Technology,
	  Cambridge, MA  02139 USA (e-mail: vgoyal@mit.edu).}}

\newtheorem{thm}{Theorem}
\newtheorem{cor}{Corollary}
\newtheorem{lem}{Lemma}
\newtheorem{prop}{Proposition}
\newtheorem{defn}{Definition}
\newtheorem{rem}{Remark}

\newcommand{\Lip}{\operatorname{Lip}}
\newcommand{\cv}{\operatorname{cv}}
\newcommand{\diam}{\operatorname{diam}}
\newcommand{\expan}{\operatorname{expan}}

\begin{document}

\maketitle

\begin{abstract}
A malleable coding scheme considers not only compression efficiency
but also the ease of alteration,
thus encouraging some form of recycling of an old compressed version in the
formation of a new one.
Malleability cost is the difficulty of synchronizing compressed versions,
and malleable codes are of particular interest when representing 
information and modifying the representation are both expensive.
We examine the trade-off between compression
efficiency and malleability cost under a malleability metric
defined with respect to a string edit distance. 
This problem introduces a metric topology to the compressed domain.
We characterize the achievable rates and malleability as the solution 
of a subgraph isomorphism problem.  This can be used to argue that
allowing conditional entropy of the edited message given the original
message to grow linearly with block length creates an exponential increase
in code length.
\end{abstract}


\newpage

\section{Introduction}
\label{sec:introduction}
\IEEEPARstart{T}{he} source coding theorem for block codes is obtained 
by calculating the number of typical source sequences and generating a set 
of labels to enumerate them.  Asymptotically
almost surely (a.a.s), only typical sequences will occur 
so it is sufficient that the set of labels be as large as the set of 
typical sequences; this yields the achievable entropy bound.  As Shannon comments,\footnote{From~\cite{Shannon1948} with
emphasis added.} ``The high probability group 
is coded in an \emph{arbitrary} one-to-one way into this set,'' 
and so in this sense there is no notion  of topology of typical sequences.  

If one is concerned with zero error rather than a.a.s.\ negligible error,
the source coding theorem for variable-length codes also yields
the entropy
as an achievable lower bound. 
In this setting, the mapping from source sequences to labels is not allowed 
to be quite as arbitrary; however, as long as an optimizing set of code lengths 
is correctly matched to source letters, there are still some arbitrary
choices in an optimal construction~\cite{Huffman1952}.  

In contrast to these well-known settings,
we investigate the mapping from the source to its compressed representation
motivated by the following problem.
Suppose that after compressing a source $X_1^n$, it is modified to 
become $Y_1^n$ according to a memoryless editing process $p_{Y|X}$.
A \emph{malleable coding} scheme preserves 
some portion of the codeword of $X_1^n$ and modifies the remainder
into a new codeword from which $Y_1^n$ may be 
decoded reliably.

There are several ways to define how one preserves some portion of the
codeword of $X_1^n$.  Here we concentrate on a \emph{malleability cost}
defined by a normalized edit distance in the compressed domain.
This is motivated by systems where the old codeword is stored in a
rewritable medium;  cost is incurred when a symbol has to be changed in
value, regardless of the location.  
Recalling the ancient practice of scraping and overwriting parchment \cite{NetzN2007},  
we call the storage medium a \emph{compressed palimpsest} and the
characterization of the trade-offs the \emph{palimpsest problem}.

A companion paper~\cite{KusumaVG2008a} focuses on a distinct problem
with a similar motivation.  There, we fix a part of the old codeword to
be recycled in creating a codeword for $Y_1^n$. 
Without loss of generality,
the fixed portion can be taken to be the beginning of the codeword,
so the new codeword is a fixed prefix followed by a new suffix.
This formulation is suitable for applications in which the update information
(new suffix)
must be transmitted through a communication channel. 
If the locations of the changed symbols were to be arbitrary,
one would need to assign a cost to the indexing of the locations.

The main result for the palimpsest problem is a graphical characterization of
achievable rates and number of editing operations.  The result involves
the solution to the error-tolerant attributed subgraph isomorphism 
problem \cite{MessmerB1998}, which is essentially a graph
embedding problem.  Although graph functionals 
such as independence number \cite{Shannon1956} and chromatic number \cite{Witsenhausen1976}\footnote{The 
chromatic number of a graph can be related to its genus (which is defined
by the topological embedding of the graph into closed, oriented surfaces \cite{GrossT1987,Tlusty2007}),
however our interest is in metric graph embedding rather than topological graph embedding.}
often arise in
the solution of information theory problems, this seems to be
the first time that the subgraph isomorphism problem 
has arisen.  Moreover, this seems to be the first treatment of 
the source code as a mapping between metric spaces.

Several of the results we obtain are pessimistic.  Unless the old source and
the new source are very strongly correlated, a large rate penalty must be
paid in order to have minimal malleability cost.  Similarly, a large
malleability cost must be incurred if the rates are required to be close
to entropy.

\subsubsection*{Outline and Preview}
The remainder of the paper is organized as follows.  In Section~\ref{sec:te},
we present a few toy examples of coding methods that
exhibit a large range of possible trade-offs.
Section~\ref{sec:background} provides additional motivation and context for our work.
Section~\ref{sec:problemstatement} then provides a formal problem statement,
and constructive coding techniques paralleling those previewed in
Section~\ref{sec:te} are developed precisely
in Section~\ref{sec:ConstPalimEx}.  

In Section~\ref{sec:sourceGraphEmbed}, graph embedding techniques are used 
to specify achievable rate--malleability points.  In particular, 
Section~\ref{sec:sourceGray} deals with Hamming distance as the editing 
cost and proposes a construction using Gray codes.  Lower bounds and 
constructive examples using letter-by-letter encoding and decoding are given.
This graph embedding approach is generalized in Section~\ref{sec:minimalchangecodes} 
to include other edit distances via generalized minimal change codes.  

While the above delay-free encoding and decoding gives optimal
results for a few special cases, we consider a more general coding
approach in Section~\ref{sec:palimpSol}, considering both variable-length 
and block codes. 
In the latter case, we 
show that the topology of typical sequences plays an important
role in our problem.  Using graph-theoretic ideas, we give
an achievability result in Theorem~\ref{thm:achBlockPalimpsest}.
Further, in Theorem~\ref{thm:expBlockPalimpsest} we argue that 
a linear reduction in malleability is at exponential cost in 
compression efficiency, consistent with the examples given
in Section~\ref{sec:te}.  This theorem is proved for ``stationary editing distributions,''
though we believe it to be true for general distributions.
In Theorem \ref{thm:palimpsestLipsz}, we give an upper bound on 
malleability cost using the Lipschitz constant of the source code mapping for general distributions.

Section \ref{sec:final} provides some final
observations on the trade-off between malleability cost and compression 
efficiency, gives some conclusions, and discusses future work.

\section{Simple Examples}
\label{sec:te}
To motivate this exposition prior to defining all quantities precisely,
we begin by giving four examples
of how one can trade off between compression efficiency and malleability.
Let $X$, $Y$, and $Z$ be binary variables with entropies $H(X)$, $H(Y)$, and $H(Z)$, respectively.
Suppose that the original observation is a word $X_1^n$.  After compressing $X_1^n$, 
the original source is modified by adding a binary sequence $Z_1^n$ with
Hamming weight $np$ to obtain a new word 
$Y_1^n = X_1^n \oplus Z_1^n$. 
Suppose the storage alphabet is also binary and that
the cost of synchronization is measured with the extended Hamming distance.
Unlike many source coding problems where 
only the cardinality of the set of codewords is used, here the alphabet itself is used
to measure malleability cost; an abstract
set of indices is not appropriate.

\begin{figure}
 \begin{center}
  \qquad
  \psfrag{a}[l][l]{\footnotesize \emph{a)}}
  \psfrag{b}[l][l]{\footnotesize \emph{b)}}
  \psfrag{c}[l][l]{\footnotesize \emph{c)}}
  \psfrag{d}[l][l]{\footnotesize \emph{d)}}
  \psfrag{a1}[][t]{\footnotesize $n$}
  \psfrag{a2}[r][r]{\footnotesize $np$}
  \psfrag{b1}[][t]{\footnotesize $nH(X)$}
  \psfrag{b2}[r][r]{\footnotesize $\frac{1}{2}nH(X)$}
  \psfrag{c1}[][t]{\footnotesize $n(H(X)+H(Z))$}
  \psfrag{c2}[r][r]{\footnotesize $nH(Z)$}
  \psfrag{d1}[][t]{\footnotesize $2^{nH(X)}$}
  \psfrag{d2}[r][r]{\footnotesize $2$}
  \psfrag{x}[l][l]{\small compressed size}
  \psfrag{y}[][t]{\small editing cost}
  \epsfig{figure=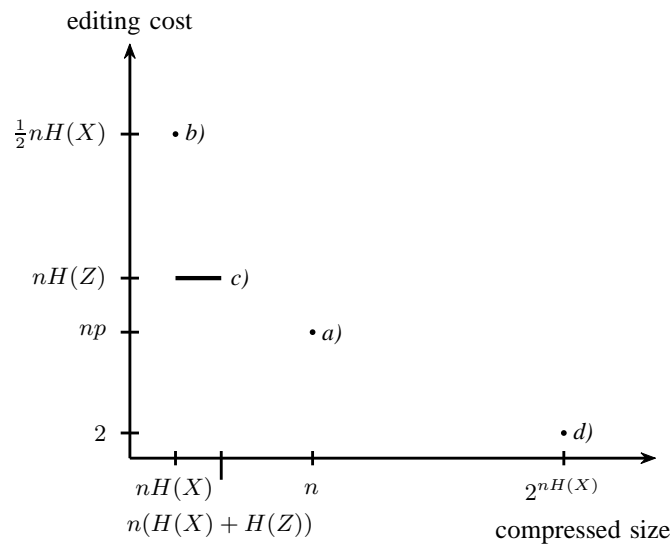,width=3in}
 \end{center}
 \caption{Qualitative representation of the four simple techniques of
     Section~\ref{sec:te}. 
     For ease of representation, it is assumed that $H(X) = H(Y)$. 
     The relative orderings of points are based on $H(Z) \ll H(X)$;
     this reflects the natural case where the editing operation
     is of low complexity relative to the original string.}
 \label{fig:preview}
\end{figure}

How might the code for $X_1^n$ and the update mechanism to allow representation
of $Y_1^n$ be designed?
The four possibilities below are summarized in Fig.~\ref{fig:preview}.

\paragraph{No compression}
We store $n$ bits for $X_1^n$.
Hence synchronizing to the new version only requires changing
the same number of bits in the code as were changed from $X_1^n$ to
$Y_1^n$; the cost is the Hamming weight of $Z_1^n$, $np$.  

\paragraph{Fully compress $X_1^n$ and $Y_1^n$}
We apply Shannon-type compression, storing
only $n H(X)$ bits for $X_1^n$.  It seems, however,
that a large portion of this old codeword will have to be changed---perhaps
about half the bits---to
become a representation for $Y_1^n$.  Compression efficiency is obtained 
at the cost of malleability.  

\paragraph{Fully compress $X_1^n$ and an increment}
Another coding strategy is to compress the change $Z_1^n$ separately and append
it to the original compression of $X_1^n$.  The new compression then
has length $n (H(X)+H(Z)) \geq n H(Y)$ bits.  The extended Hamming malleability cost
is $n H(Z)$ bits.

\paragraph{Completely favor malleability over compression}
Interestingly, there is a method that dramatically trades compression
efficiency for malleability.\footnote{Due to Robert 
G. Gallager.}  The source $X_1^n$ is encoded with
$2^{nH(X)}$ bits, using an indicator function to denote which
of its typical sequences was observed.  The same strategy is
used to encode $Y_1^n$, using $2^{nH(Y)}$ bits.
Then synchronization requires
changing only two bits when $X_1^n$ and $Y_1^n$ are different.  

Our purpose is to study the limits of this interesting trade-off between
compression efficiency and malleability.
We will do so using formalized performance metrics after a bit more
background.

\section{Background}
\label{sec:background}
Our study of malleable compression is motivated by information storage systems that
store documents which are updated often.  In such systems, the storage costs include
not only the average length of the coded signal, but also the costs in updating.  
We describe these systems and also discuss an information storage system in synthetic 
biology, where the editing costs are much more significant and restrictive 
than in optical or 
magnetic systems. 

\subsection{Version Management}
Consider the installation of a security patch to an operating system, the update of 
a text document after proofreading, the storage of a computer file backup 
system after a day's work, or a second email that corrects the location of a seminar
yet also reproduces the entire seminar abstract.  In all of these settings and 
numerous others, separate data streams may be generated, but the contents differ only 
slightly \cite{BobbarjungJD2006,PolicroniadesP2004,BurnsSL2003,SuelM2003}.  Moreover, in 
these applications, old versions of the stream need not be preserved.  Particularly 
for devices such as mobile telephones, where memory size and
energy are severely constrained, but for any storage system, it is advisable to 
reduce the space taken by data and also to reduce the energy required to insert, 
delete, and modify stored data.  In certain applications, in-place reconstruction is
desired \cite{BurnsSL2003}, necessitating the use of instantaneous source codes.  

Recursive estimation and control also require temporarily storing state estimates and 
updating them at each time step.  Thus such problems also suggest themselves
as application areas for malleable coding.  Note that the application
of malleable codes would determine how information storage is carried out, not
what information is stored and what information is dissipated \cite{MitterN2005}.

In the scenarios discussed, new versions will be 
correlated with old versions, not independent as assumed in previous studies of
write-efficient memories \cite{AhlswedeZ1989,AhlswedeZ1994}.  That is, we envision
scenarios which involve updating {\small \tt Archimedes of Siracusa} with {\small \tt Archimedes of Syracuse}  
(Levenshtein distance $2$) rather than updating with {\small \tt Jesus of Nazareth}  
(Levenshtein distance $15$), though the results will apply to the entire gamut of scenarios.

There is also another difference between the problems we formulate and prior work 
on write-efficient memories.  In write-efficient memories, the encoder can look to see 
what is already stored in the memory before deciding the codeword for the update.
An information pattern even more extensive than for write-efficient memories
was discussed in \cite{RamprasadSH1999c}.
We require the code to be determined before the encoding process is carried out.  Such an 
information pattern would arise naturally in remote file synchronization \cite{SuelM2003}.  

Once the codeword of the new version is 
determined (without access to the realized compressed old version), there may be 
settings where the differences between the two must be determined in a distributed
fashion.
For a good malleable code, the
old and new codewords will be strongly correlated.  Thus, protocols for distributed
reconciliation of correlated strings may be used \cite{Orlitsky1993,MinskyTZ2003,SuelM2003}.

\subsection{Genetic Coding}
With recent advances in biotechnology \cite{GarfinkelEEF2007}, the storage 
of artificial messages in DNA strings seems like a real possibility, rather than 
just a laboratory pipe dream \cite{WongWF2003}.  Thus the storage of messages in 
the DNA of living organisms as a long-lasting, high-density data storage 
medium provides another motivating application for malleable coding.
Note that although minimum change codes, as we will develop for the palimpsest problem,
have been suggested as an explanation for the genetic code through the optimization 
approach to biology \cite{Swanson1984}, here we are concerned with synthetic biology.  

As in magnetic or optical storage and perhaps more so, it is desirable to compress 
information for storage.  For a palimpsest system, one would use site-directed 
mutagenesis \cite[Ch. 7]{PrimroseTO2001} to perform editing of stored codewords 
whereas for the formulation of malleable coding in~\cite{KusumaVG2008a}, molecular biology cloning techniques using 
restriction enzymes, oligonucleotide synthesis or polymerase chain reaction (PCR), 
and ligation \cite[Ch. 3]{PrimroseTO2001} would be used.  
In site-directed mutagenesis when multiple changes cannot be made using a single primer,
the cost of a single insertion, deletion, or 
substitution is approximately the same and is additive with respect to the number of 
edits.  Using restriction enzyme methods with oligonucleotide synthesis, however, the cost is related to the length 
of the new segment that must be synthesized to replace the old segment.  
Thus the biotechnical editing costs correspond exactly to the costs defined in
the present paper and in~\cite{KusumaVG2008a}.
Unlike magnetic or optical storage, insertion and deletion are natural operations 
in DNA information storage, thereby 
allowing variable-length codes to be easily edited.  Incidentally,
insertion and deletion is also possible in neural information 
storage through modification of neuronal arbors \cite{VarshneySC2006}. 

\section{Problem Statement}
\label{sec:problemstatement}
After a few requisite definitions,
we will provide a formal statement of the palimpsest problem,
which takes editing costs as well as rate costs into account.  

The symbols of the storage medium are drawn from the finite alphabet
$\mathcal{V}$.
Note that unlike most source coding problems,
the alphabet itself will be used,
not just the cardinality of sequences drawn from this alphabet.
Also, it is natural to measure all rates in numbers of symbols from
$\mathcal{V}$.  This is analogous to using base-$|\mathcal{V}|$ logarithms
in place of base-2 logarithms, and all logarithms should be interpreted
as such.

We require the notion of an
edit distance \cite{Cormode2003} on $\mathcal{V}^{*}$,
the set of all finite sequences of elements of $\mathcal{V}$.
\begin{defn}
An \emph{edit distance}, $d(\cdot,\cdot)$, is a function from $\mathcal{V}^{*}\times\mathcal{V}^{*}$ 
to $[0,\infty)$, defined by a set of edit operations.  The edit operations are a symmetric
relation on $\mathcal{V}^{*}\times\mathcal{V}^{*}$.  The edit distance between
$a \in \mathcal{V}^{*}$ and $b \in \mathcal{V}^{*}$ is $0$ if $a = b$ and is the minimum 
number of edit operations needed to transform $a$ into $b$ otherwise.
\end{defn}

An example of an edit distance is the Levenshtein distance, which is constructed from 
insertion, deletion, and substitution operations.  It can be noted that $(\mathcal{V}^*,d)$
is a finite metric space (see Appendix~\ref{app:prop-edit}).

Now we can formally define our coding problem. 
We define the variable-length and block coding versions together,
drawing distinctions only where necessary.
Symbols are reused so as to conserve notation. 
It should be clear from context whether we are discussing
variable-length or block coding.

Let $\{(X_i,Y_i)\}_{i=1}^{\infty}$ be a sequence of independent drawings
of a pair of random variables $(X,Y)$, $X \in \mathcal{W}$, $Y \in \mathcal{W}$, where
$\mathcal{W}$ is a finite set and $p_{X,Y}(x,y) = \Pr[X = x, Y=y]$.
The marginal distributions are
\[
p_X(x) = \sum_{y\in\mathcal{W}} p_{X,Y}(x,y)
\]
and
\[
p_Y(y) = \sum_{x\in\mathcal{W}} p_{X,Y}(x,y) \mbox{.}
\]
When the random variable is clear 
from context, we write $p_X(x)$ as $p(x)$ and so on.

A modification channel 
\[
p_{Y|X}(y|x) = \frac{p(x,y)}{p(x)}
\]
relates the two marginal distributions.  If the joint distribution is such 
that the marginals are equal, the modification channel is said to perform 
\emph{stationary editing}. 

\subsubsection*{Variable-length Codes}
A variable-length encoder with block length $n$ is a mapping
\[
f_E: \mathcal{W}^n \to \mathcal{V}^{*}\mbox{,}
\]
and the corresponding decoder with block length $n$ is
\[
f_D: \mathcal{V}^{*} \to \mathcal{W}^n \mbox{.}
\]
The encoder and decoder define a variable-length palimpsest code.
The encoder and decoder pair is required to be instantaneous, in the 
sense that the encoding may be parsed as a succession of codewords.

A (variable-length) encoder-decoder with block length $n$ is applied as follows.
Let
\[
(A,B) = (f_E(X_1^n),f_E(Y_1^n)) \mbox{,}
\]
inducing random variables $A$ and $B$ that are drawn from the alphabet
$\mathcal{V}^{*}$.  Also let
\[
(\hat{X}_1^n,\hat{Y}_1^n) = (f_D(A),f_D(B)) \mbox{.}
\]

\subsubsection*{Block Codes}
A block encoder for $X$ with parameters $(n,K)$ is a mapping
\[
f_E^{(X)}: \mathcal{W}^n \to \mathcal{V}^{nK}\mbox{,}
\]
and a block encoder for $Y$ with parameters $(n,L)$ is a mapping
\[
f_E^{(Y)}: \mathcal{W}^n \to \mathcal{V}^{nL} \mbox{.}
\]
Given these encoders, a common decoder with parameter $n$ is
\[
f_D: \mathcal{V}^{*} \to \mathcal{W}^n \mbox{.}
\]
The encoders and decoder define a block palimpsest code. 
Since there is a common decoder, the two codes should be in the same format.

\begin{figure}
  \centering
  \includegraphics[width=3.0in]{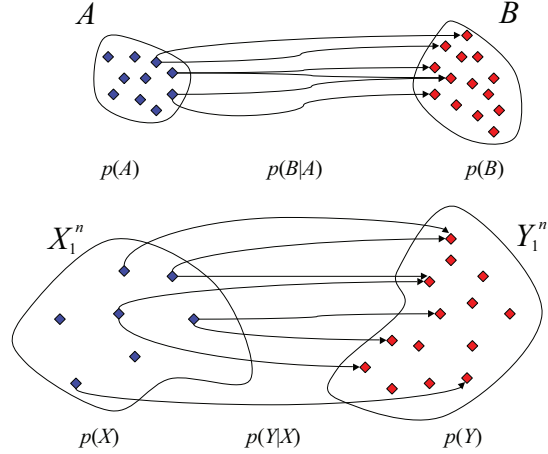}
    \caption{Distributions in representation space induced by distributions in source space.}
  \label{fig:combstab_gen}
\end{figure}

A (block) encoder-decoder with parameters $(n,K,L)$ is applied as follows.  Let
\[
(A,B) = (f_E^{(X)}(X_1^n),f_E^{(Y)}(Y_1^n)) \mbox{,}
\]
inducing random variables $A \in \mathcal{V}^{nK}$ and $B \in \mathcal{V}^{nL}$.
The mappings are depicted in Fig.~\ref{fig:combstab_gen}.  
Also let 
\[
(\hat{X}_1^n,\hat{Y}_1^n) = (f_D(A),f_D(B)) \mbox{.}
\]

For both variable-length and block coding,
we can define the error rate as
\[
\Delta = \max(\Delta_X,\Delta_Y) \mbox{,}
\]
where
\[
\Delta_X = \Pr[X_1^n \neq \hat{X}_1^n]
\]
and
\[
\Delta_Y = \Pr[Y_1^n \neq \hat{Y}_1^n] \mbox{.}
\]

Natural (and completely conventional)
performance indices for the code are the per-letter average 
lengths of the codewords
\[
K = \frac{1}{n} E\left[ \ell(A) \right] \mbox{,}
\]
and
\[
L = \frac{1}{n} E\left[ \ell(B) \right] \mbox{,}
\]
where $\ell(\cdot)$ denotes the length of a sequence in $\mathcal{V}^{*}$.
(In the block coding case, $A$ has a fixed length of $nK$ letters from
the alphabet $\mathcal{V}$, so there is no contradiction in using
the previously-defined symbol $K$.  Similarly for $L$.)

The final performance measure captures our novel concern with the
cost of changing the compressed version.
The malleability cost is the
expected per-source-letter edit distance between the codes:
\[
M = \frac{1}{n} E\left[ d(A,B) \right] \mbox{.}
\]

\begin{defn}
\label{defn:P_V}
Given a source $p(X,Y)$ and an edit distance $d$, a triple $(K_0,L_0,M_0)$ 
is said to be \emph{achievable} for the variable-length palimpsest problem if,
for arbitrary $\epsilon > 0$, 
there exists (for $n$ sufficiently large)
a variable-length palimpsest code with error rate $\Delta = 0$,
average codeword lengths $K \le K_0 + \epsilon$, 
$L \le L_0 + \epsilon$, and malleability $M \le M_0 + \epsilon$.  
\end{defn}

\begin{defn}
\label{defn:P_B}
Given a source $p(X,Y)$ and an edit distance $d$, a triple $(K_0,L_0,M_0)$ 
is said to be \emph{achievable} for the block palimpsest problem if,
for arbitrary $\epsilon > 0$, 
there exists (for $n$ sufficiently large)
a block palimpsest code with error rate $\Delta < \epsilon$,
average codeword lengths $K \le K_0 + \epsilon$, 
$L \le L_0 + \epsilon$, and malleability $M \le M_0 + \epsilon$.  
\end{defn}

For the variable-length palimpsest problem,
the set of achievable rate--malleability triples is denoted by $\mathfrak{P}_V$;
for the block version, the corresponding set is denoted by $\mathfrak{P}_B$.
It will be our purpose to characterize
$\mathfrak{P}_V$ and $\mathfrak{P}_B$ as much as possible.

It follows from the definition that $\mathfrak{P}_V$ and $\mathfrak{P}_B$
are closed subsets of $\mathbb{R}^3$ and have the property
that if $(K_0,L_0,M_0) \in \mathfrak{P}$, then 
$(K_0+\epsilon_1,L_0+\epsilon_2,M_0+\epsilon_3) \in \mathfrak{P}$ for any
$\epsilon_i \ge 0$, $i = 1,2,3$. 
Consequently, $\mathfrak{P}_V$ and $\mathfrak{P}_B$ are completely defined by
their lower boundaries, which too are closed.  

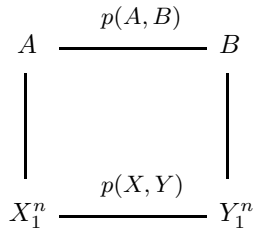
\begin{figure}
  \centering
  \setlength{\unitlength}{0.22in}
  \begin{picture}(6.0,5.0)(0.0,0.0)
  \thicklines
  \put(0.0,0.0){\small $X_1^n$}
  \put(5.0,0.0){\small $Y_1^n$}
  \put(0.2,4.0){\small $A$}
  \put(5.0,4.0){\small $B$}
  \put(1.2,0.1){\line(1,0){3.5}} \put(1.2,4.1){\line(1,0){3.5}}
  \put(0.4,1.0){\line(0,1){2.5}} \put(5.2,1.0){\line(0,1){2.5}}
  \put(2.2,0.7){\footnotesize $p(X,Y)$}
  \put(2.2,4.7){\footnotesize $p(A,B)$}
  \end{picture}
   \caption{Commutative diagram for the palimpsest problem.}
  \label{fig:cs_commut}
\end{figure}

Both versions of the palimpsest problem can be viewed using the diagram
in Fig.~\ref{fig:cs_commut}.  Given $p(X,Y)$ 
and thus $p(X)$, $p(Y)$, and $p(Y|X)$, the malleability constraint 
defines what is achievable in terms of $p(A,B)$ with the additional constraints 
that there must be maps between $X_1^n$ and $A$, and between $Y_1^n$ and $B$, which
allow for lossless or near lossless compression.  An alternative formulation as
the mapping between two metric spaces $\mathcal{W}^n$ and $\mathcal{V}^*$ is also possible.  

\section{Constructive Palimpsest Examples}
\label{sec:ConstPalimEx}
Having formulated the palimpsest problem in Section~\ref{sec:problemstatement},
we present some examples of what can be achieved.  These examples revisit Section~\ref{sec:te}.
New examples given in Section~\ref{sec:sourceGraphEmbed} will inspire general statements.

\subsection{Source Coding with No Compression}
The simplest compression scheme is one that simply copies the source sequences
to the storage medium.  This is only possible when 
$\mathcal{W} = \mathcal{V}$. 
When $\mathcal{W} \neq \mathcal{V}$, zero-error coding without compression
is possible with block lengths larger than 1,
as in converting hexadecimal digits to binary digits or vice versa.
The flexibility in such a mapping can be exploited.  If the shortest possible blocking is used and $l$ 
is the least common multiple of $|\mathcal{V}|$ and $|\mathcal{W}|$, then there are 
$l!$ valid mappings.  For the moment, we ignore the gains to be had by exploiting
this flexibility and focus on the $\mathcal{W} = \mathcal{V}$ case, with block 
length $n = 1$.  

Taking $A = X$ and $B = Y$, it follows immediately that 
$K = 1$ and $L = 1$.\footnote{Remember that rates $K$ and $L$ are
measured in letters from $\mathcal{V}$, not in bits.}  It also follows
that the malleability cost is $M = E[d(X,Y)]$.  If we take 
the edit distance to be the Hamming distance, then $M = \Pr[X\neq Y]$.  
Thus the triple $(K,L,M) = (1,1,\Pr[X\neq Y])$
is achievable by no compression for any source distribution $p(X,Y)$ under Hamming
edit distance.

\subsection{Ignore Malleability}
Consider what happens when the malleability parameter is ignored and 
the rates for the variable-length encoder are optimized.  We will improve
rate performance and hopefully not worsen malleability too much.

If the updating process $p_{Y|X}$ is stationary, then
a common instantaneous code may be used to asymptotically achieve 
$K = H(X)$ and $L = H(Y)$.  Picking a single code for different sources 
has been well-studied in the source coding literature, starting with 
\cite{Gilbert1971}.  If a single source code is used for a collection of
distributions, the rate loss over the entropy lower bound is termed the redundancy 
\cite{Davisson1973}.  As shown by Gilbert, 
if Huffman or Shannon codes are used, this redundancy is the relative entropy 
between the source and the random variable used to design the code.  

Restricting to such instantaneous codes, if the palimpsest code is designed for either 
$p(x)$ or for $p(y)$, the incurred redundancies are the relative entropies 
\[
D(p_X\|p_Y) = \sum\limits_{x\in\mathcal{W}}p(x) \log \frac{p(x)}{p(y)}
\]
or
\[
D(p_Y\|p_X) = \sum\limits_{y\in\mathcal{W}}p(y) \log \frac{p(y)}{p(x)}
\]
respectively.  These lead to horizontal and vertical portions of a lower bound 
for $\mathfrak{P}_V$ in the $K$--$L$ plane.

An intermediary portion of this lower bound, between the vertical 
and horizontal portions, is determined by finding a random variable $Z$ that is between 
$X$ and $Y$ and designing a code for it.  We want to choose some ``tilted'' 
distribution, $p_Z$, on the geodesic between the two distributions $p_X$ and $p_Y$.  

\begin{figure}
  \begin{center}
   \quad
   \psfrag{A}[t][t]{\scriptsize $H(X)$}
   \psfrag{B}[r][r]{\scriptsize $H(Y)$}
   \psfrag{C}[t][t]{\scriptsize $H(X) + D(X \| Y)$}
   \psfrag{D}[r][r]{\shortstack{\scriptsize $H(Y) +$ \\
                                \scriptsize $D(Y \| X)$}}
   \psfrag{K}[l][l]{$K$}
   \psfrag{L}[b][]{$L$}
   \epsfig{figure=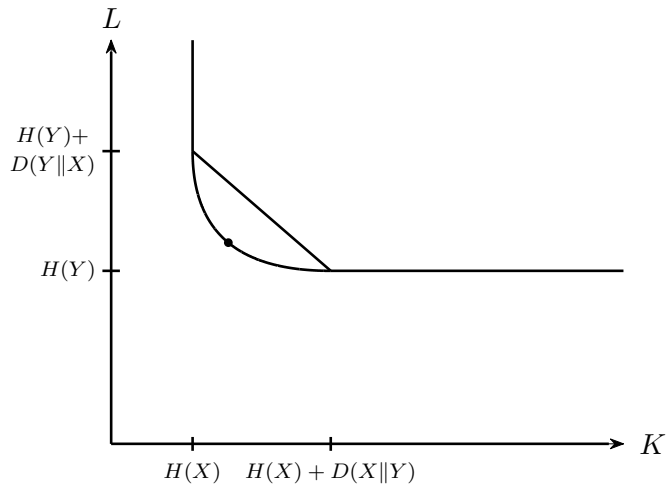,width=3in}
  \end{center}
  \caption{$K$--$L$ region achievable using instantaneous codes 
	for sources related through non-stationary editing.
	The marked point is when rate loss for both versions is 
	balanced.  The diagonal line segment shows the suboptimal 
	strategy of time-sharing.
  }
  \label{fig:differentmarginals}
\end{figure}

If $p_Z = \tfrac{1}{2}p_X + \tfrac{1}{2}p_Y$, then $D(p_X\|p_Z) + D(p_Y\|p_Z)$ is called the 
capacitory discrimination \cite{Topsoe2000}.  The rate loss in the balanced rate loss case, 
$D(p_{X}\|p_{Z})$ when $D(p_{X}\|p_{Z}) = D(p_{Y}\|p_{Z})$, has a closed form expression 
\cite{SinanovicJ2007}.  The distribution $p_Z$ used to achieve it is halfway (in the asymmetric 
sense of $Y$ after $X$) along the geodesic that connects the two distributions.  The distance
along the geodesic may be parameterized by
\[
t = \frac{D(p_Y\|p_X)}{D(p_Y\|p_X) + D(p_X\|p_Y)} \mbox{.}
\]
The resulting rate loss for $Z_t$ is
\[
D(p_X\|p_{Z_t}) = R(p_X,p_Y) + \log\mu(t) \mbox{,}
\]
where $R(p_X,p_Y)$ is defined through
\[
\frac{1}{R(p_X,p_Y)} = \frac{1}{D(p_X\|p_Y)} + \frac{1}{D(p_Y\|p_X)} \mbox{,}
\]
and
\[
\mu(t) = \sum\limits_{x\in\mathcal{W}}p_X^{1-t}(x)p_Y^{t}(x) \mbox{.}
\]
Notice that due to the asymmetry of the relative entropy, this is different than the 
Chernoff information.  In general, the connecting 
portion between the horizontal and vertical parts of the lower bound is curved below the 
time-sharing line, determined by the relative entropies $D(p_X\|p_Z)$ and $D(p_Y\|p_Z)$ for a $Z$
that is along the geodesic connecting the two distributions.  Fig.~\ref{fig:differentmarginals}
shows an example of this achievable lower bound.  

If the restriction to instantaneous codes is removed, then 
there are several kinds of universal source codes that achieve the $K \ge H(X)$ 
and $L \ge H(Y)$ bounds simultaneously \cite{Davisson1973,DavissonMPW1981}, however instantaneous
codes are required by the palimpsest problem statement.  
These results say nothing 
about $M$, they only deal with $K$ and $L$.

To say something about
$M$, one can show that the average starting overlap is rather small \cite{GacsK1973}.  
Since optimal source codes produce equiprobable outputs \cite{VisweswariahKV1998}, 
one might hope that computing $M$ is a matter of measuring 
the expected edit distance between two random equiprobable sequences \cite{FuS1997},
but optimizing the dependence between these two sequences is actually the problem to be solved.  

\subsection{Source Coding with Incremental Compression}
One might compress the original source using an 
optimal source code, thereby achieving the $K \ge H(X)$ lower bound with 
equality.  Then one may produce an optimal source code for the innovation 
separately, with rate $H(Y|X)$.  Thus the new version would be represented
by concatenating the two pieces, with $L = H(X) + H(Y|X) = H(X,Y)$.  
Under extended Hamming edit distance, the difference 
between the original source code and the new version which has a new piece
concatenated would be $M = H(Y|X)$. 

Separate compression of the innovation has the advantage that $X_1^n$ can be 
recovered from $B$, however this was not a requirement in the problem formulation
and is thus wasteful.  Such a coding scheme is useful in differential encoding for
version management systems where all versions should be recoverable.  Results
would basically follow from the chain rule of entropy \cite{Korner1971} or 
from successive refinability for a lossy version of the problem \cite{EquitzC1991}.

\subsection{Source Coding with Pulse-Position Modulation}
\label{sec:PPM}
Another coding strategy is to significantly back off 
from achieving good rate performance so as to achieve very good malleability.
In particular, we describe a compression scheme that requires only $2$ substitution
edits for any modification to the source, and so the value of $M$ 
achieved goes to $0$ asymptotically.

We represent any of the possible $|\mathcal{W}|^n$ sequences that can occur 
as $X_1^n$ or $Y_1^n$ by a pulse-position modulation scheme.  In particular,
we use only two letters from $\mathcal{V}$, which we call $0$ and $1$ without loss
of generality.
The codebook is the set of binary sequences of length $|\mathcal{W}|^n$ with Hamming weight $1$.
Each possible source 
sequence is assigned to a distinct codebook entry, thus making $\Delta = 0$.  Now modifying any sequence
to any other sequence entails changing a single $0$ to a $1$ and a single $1$ to a $0$.
Computing the performance criteria, we get that $K = L = \tfrac{1}{n}|\mathcal{W}|^n$,
and so are paying an exponential rate penalty over simply enumerating the source 
sequences.  The payoff is that $M = \frac{2}{n}\Pr[X_1^n \neq Y_1^n]$.  This is true universally,
even if $X$ and $Y$ are independent.  Note that if a.a.s.\ no error is desired, 
then only typical source sequences need to have codewords assigned to them, and so 
$K = L = \tfrac{1}{n}2^{n\max(H(X),H(Y))}$ (where $H(X)$ and $H(Y)$ are here in bits)
has the same effect on $M$.

Pulse-position modulation is also a possible scheme for achieving channel capacity 
per unit cost \cite{Verdu1990}, where an exponential spectral efficiency penalty is
paid in order to have very low power.

\section{Source Coding with Graph Embedding}
\label{sec:sourceGraphEmbed}
Before constructing an example, let us develop some lower bounds for arbitrary
sources $p(X,Y)$.  From the source coding theorems, it follows that $K \ge H(X)$
and $L \ge H(Y)$.  We observe that since distinct codewords must have an 
edit distance of at least one, we can lower bound $M$ by assuming that distance
$1$ is achieved for all codewords.  Then the edit distance is simply the
probability of error for uncoded transmission of $p(x)$ through the channel
$p(y|x)$, since each error gives edit distance $1$ and each correct reception
gives edit distance $0$.  Thus for $n = 1$, 
$M\ge\sum_{x\in\mathcal{W}}\sum_{y\in\mathcal{W}: y \neq x} p(x,y)$.  
For larger $n$, the bound is similarly derived to be
\begin{equation}
\label{eqn:GraphEmbedLowerbound}
M\ge\frac{1}{n}\sum\limits_{x_1^n\in\mathcal{W}^n}\sum\limits_{y_1^n\in\mathcal{W}^n: y_1^n \neq x_1^n} p(x_1^n,y_1^n) \mbox{.}
\end{equation}
A weaker, simplified version of the bound is $M \ge \tfrac{1}{n}$.  As will be evident
in the sequel, this weaker bound is related to Lipschitz constants for the mapping from 
the source space to the representation space.

\subsection{Graph Embedding using Gray Codes}
\label{sec:sourceGray}
Now we construct an example that simultaneously achieves the rate lower bounds
and the malleability lower bound (\ref{eqn:GraphEmbedLowerbound}).
Consider a memoryless source $p(x)$ with alphabet 
$\mathcal{W} = \{${\fransdvng k}, {\fransdvng K}, {\fransdvng G}, {\fransdvng g}, 
{\fransdvng j}, {\fransdvng J}, {\fransdvng C}, {\fransdvng c}$\}$, 
such that each letter is drawn equiprobably.\footnote{The scholar of linguistics and 
coding theory will notice the relevance of 
the order in which the alphabet is written \cite{VarshneyG2006b}.}
Then the original version of the source has entropy $3$ bits.  
Consider the relationship between $X$ and $Y$  given by a noisy 
typewriter channel, with channel transmission matrix 
\begin{equation}
\label{eqn:NoisyTypewriter}
p(y|x) = \left[ {\begin{array}{*{20}c}
   \tfrac{1}{2} & \tfrac{1}{4} & 0 & 0 & 0 & 0 & 0 & \tfrac{1}{4} \vspace{.5mm}  \\
   \tfrac{1}{4} & \tfrac{1}{2} & \tfrac{1}{4} & 0 & 0 & 0 & 0 & 0 \vspace{.5mm} \\
   0 & \tfrac{1}{4} & \tfrac{1}{2} & \tfrac{1}{4} & 0 & 0 & 0 & 0 \vspace{.5mm}  \\
   0 & 0 & \tfrac{1}{4} & \tfrac{1}{2} & \tfrac{1}{4} & 0 & 0 & 0 \vspace{.5mm} \\
   0 & 0 & 0 & \tfrac{1}{4} & \tfrac{1}{2} & \tfrac{1}{4} & 0 & 0 \vspace{.5mm} \\
   0 & 0 & 0 & 0 & \tfrac{1}{4} & \tfrac{1}{2} & \tfrac{1}{4} & 0 \vspace{.5mm} \\
   0 & 0 & 0 & 0 & 0 & \tfrac{1}{4} & \tfrac{1}{2} & \tfrac{1}{4} \vspace{.5mm} \\
   \tfrac{1}{4} & 0 & 0 & 0 & 0 & 0 & \tfrac{1}{4} & \tfrac{1}{2} \vspace{.5mm} \\
\end{array}} \right] \mbox{.}
\end{equation}
Evidently, the bound on $M$ is $1/2$ for $n=1$, found by performing the 
summation in (\ref{eqn:GraphEmbedLowerbound}).  Moreover, the marginal distribution of $y$ is also equiprobable from the 
alphabet $\mathcal{W}$, which gives the entropy bound on $L$ to be $3$ bits.

\begin{figure}
  \centering
  \includegraphics[width=2in]{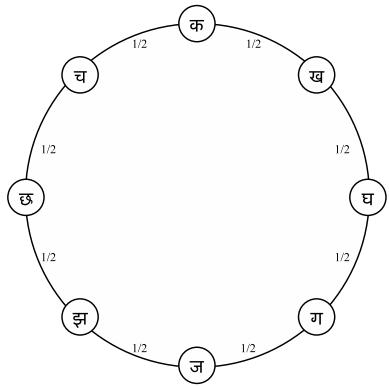}
  \caption{Weighted adjacency graph for noisy typewriter channel
           (\ref{eqn:NoisyTypewriter}).}
  \label{fig:wAdjGraph}
\end{figure}

Take $\mathcal{V}$ to be $\{0,1\}$.  Now we develop a binary 
encoding scheme that has performance coincident with the established inner 
bounds, using graph embedding methods.  We can 
draw a graph where the vertices are the symbols and the edges are labeled 
with the associated probabilities of transition;  the weighted 
directed edges are combined into weighted undirected edges in some suitable way.  The result 
is a weighted adjacency graph, a weighted version of the adjacency 
graphs in \cite{Shannon1956, Witsenhausen1976}, shown in 
Fig.~\ref{fig:wAdjGraph}.  

\begin{figure}
  \centering
  \includegraphics[width=1.7in]{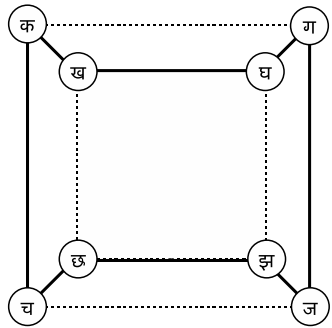}
  \caption{Weighted adjacency graph for noisy typewriter channel embedded in 3-dimensional hypercube.  Thick lines represent edges that are used in the embedding.  Dotted lines represent edges in the hypercube that are unused in the embedding.}
  \label{fig:wAdjGraph_hypercube}
\end{figure}

\begin{figure}
  \centering
  \includegraphics[width=1.7in]{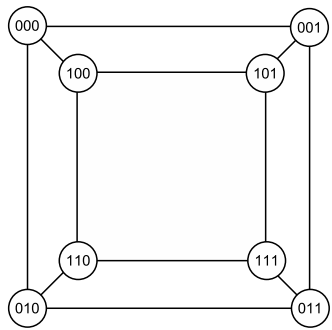}
  \caption{Hypercube graph labeled with binary reflected Gray code.}
  \label{fig:brGraycode}
\end{figure}

Suppose that the edit distance is the Hamming distance.   
Now we try to embed this adjacency graph into a hypercube of a given size.  
Since we want the average code length to be small, we first consider 
the hypercube of size $3$.  The adjacency graph is exactly embeddable into  
the hypercube, as shown in Fig.~\ref{fig:wAdjGraph_hypercube}.  If it 
were not exactly embeddable, some of the low weight edges might have 
to be broken.  As an alternative to the edge weights being determined 
from the transition matrix, the edge weights may be determined through 
a joint typicality measure (as in the message graph in 
\cite{PradhanCR2006} and in Section \ref{sec:palimpSol}).  After 
we complete the embedding into the hypercube, we use the binary 
reflected Gray code (see e.g. \cite{AgrellLSO2004} for a description) 
to assign codewords through correspondence.  The binary reflected Gray code-labeled
hypercube is shown in Fig.~\ref{fig:brGraycode}.   

Clearly the error rate for this scheme is $\Delta = 0$,
since the code is lossless.  Since all codewords are of 
length $3$, clearly $K = L = 3$.  To compute $M$, notice that any source 
symbol is perturbed to one of its neighbors with probability $1/2$.  Further 
notice that the Hamming distance between neighbors in the hypercube is $1$.  
Thus $M = 1/2$.  We have seen that this encoding scheme 
achieves the entropy bounds $H(X)$ and $H(Y)$.  It also achieves the $n=1$ lower
bound for $M$ 
and is thus optimal for $n=1$.  We can further drive $M$ down 
by increasing the block length.  As shown in the following proposition, if a graph
is embeddable in another graph and we take Cartesian products of each with 
itself, then the resulting graphs obey the same embedding relationship.

\begin{defn}
\label{def:embeddable}
Consider two graphs $G$ and $H$ with vertices $V(G)$ and $V(H)$ and edges 
$E(G)$ and $E(H)$, respectively.  Then $G$ is said to be \emph{embeddable}
into $H$ if $H$ has a subgraph isomorphic to $G$.  That is, there is an 
injective map $\phi: V(G) \to V(H)$ such that $(u,v) \in E(G)$ 
implies $(\phi(u),\phi(v)) \in E(H)$.  This is denoted as $G \leadsto H$.
\end{defn}

\begin{defn}
Consider two graphs $G_1$ and $G_2$ with vertices $V(G_1)$ and $V(G_2)$ and edges 
$E(G_1)$ and $E(G_2)$, respectively.  Then the \emph{Cartesian product} of $G_1$
and $G_2$, denoted $G_1 \times G_2$ is a graph with vertex set $V(G_1) \times V(G_2)$
and for vertices $u = (u_1,u_2)$ and $v = (v_1,v_2)$, $(u,v) \in E(G_1 \times G_2)$
when ($u_1=v_1$ and $(u_2,v_2) \in E(G_2)$) or ($u_2=v_2$ and $(u_1,v_1) \in E(G_1)$).
\end{defn}

\begin{prop}
\label{prop:cartesian-product}
If $G_1 \leadsto H_1$ and $G_2 \leadsto H_2$, then $G_1\times G_2 \leadsto H_1\times H_2$.
A special case is that $G \leadsto H$ implies $G\times G \leadsto H\times H$.
\end{prop}
\begin{IEEEproof}
See Appendix~\ref{app:prop-cartesian-product}.
\end{IEEEproof}
\begin{cor}
Let $G^n$ denote the $n$-fold Cartesian product of $G$ and $H^n$ the $n$-fold 
Cartesian product of $H$.  If $G \leadsto H$, then $G^n \leadsto H^n$ for 
$n = 1, 2, \ldots$.
\end{cor}
\begin{IEEEproof}
By induction.
\end{IEEEproof}

Returning to our example,
since the embedding relation is true for $n = 1$, it is also true for 
$n = 2,\ldots$, so we can embed
$n$-fold Cartesian products of the adjacency graph into $n$-fold Cartesian
products of the hypercube.  Such a scheme would achieve rates of $K = 3$ 
bits and $L = 3$ bits.  It would also achieve $M$ of 
$\tfrac{1}{n} \Pr[X_1^n \neq Y_1^n]$ since the Cartesian product of the 
adjacency graph represents exactly edit costs of $1$. 
For each $n$, this matches the lower 
bounds given in (\ref{eqn:GraphEmbedLowerbound}), and is thus optimal. 
Furthermore, asymptotically in $n$, the triple $(K,L,M) = (3,3,0)$ is achievable. 

One may observe that embeddability into a graph
where graph distance corresponds to edit distance seems to be sufficient 
to guarantee good performance; we will explore this in detail in the sequel.
But first, we present a similar but more challenging situation as a contrast
to the ``best of all worlds'' performance we have just seen.

With the source alphabet, representation alphabet,
and distribution of $X$ remaining the same,
let us suppose that the relationship between $X$ and $Y$ is given by 
\begin{equation}
\label{eqn:editProcess2}
p(y|x) = \left[ {\begin{array}{*{20}c}
\tfrac{2}{5} & \tfrac{1}{5} & \tfrac{1}{20} & \tfrac{1}{5} & 0 & 0 & 0 & \tfrac{3}{20} \vspace{.5mm}  \\
\tfrac{1}{5} & \tfrac{3}{5} & 0 & 0 & 0 & 0 & \tfrac{1}{5} & 0 \vspace{.5mm} \\
\tfrac{1}{20} & 0 & \tfrac{3}{5} & 0 & 0 & \tfrac{7}{20} & 0 & 0 \vspace{.5mm}\\
\tfrac{1}{5} & 0 & 0 & \tfrac{3}{5} & \tfrac{1}{5} & 0 & 0 & 0 \vspace{.5mm} \\
0 & 0 & 0 & \tfrac{1}{5} & \tfrac{3}{5} & 0 & 0 & \tfrac{1}{5} \vspace{.5mm} \\
0 & 0 & \tfrac{7}{20} & 0 & 0 & \tfrac{3}{5} & 0 & \tfrac{1}{20} \vspace{.5mm}\\
0 & \tfrac{1}{5} & 0 & 0 & 0 & 0 & \tfrac{3}{5} & \tfrac{1}{5} \vspace{.5mm} \\
\tfrac{3}{20} & 0 & 0 & 0 & \tfrac{1}{5} & \tfrac{1}{20} & \tfrac{1}{5} & \tfrac{2}{5} \vspace{.5mm} \\
\end{array}} \right] \mbox{.}
\end{equation}
One can verify that, like (\ref{eqn:NoisyTypewriter}), this is a stationary
editing process.
Thus, the rate bounds are unchanged at $K \geq 3$ and $L \geq 3$.
Also, evaluation of (\ref{eqn:GraphEmbedLowerbound}) yields the bound
$M \geq \frac{9}{20}$ for block size $n=1$.
We will presently see that the three lower bounds cannot be achieved
simultaneously, and we will determine the best values of $(K,L,M)$
for $n=1$.

\begin{figure}
  \centering
  \includegraphics[width=1.7in]{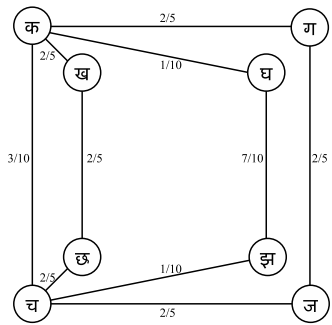}
  \caption{Weighted adjacency graph for stationary editing process
           (\ref{eqn:editProcess2}).}
  \label{fig:wAdjGraph2}
\end{figure}

The weighted adjacency graph corresponding to the new editing process
is depicted in Fig.~\ref{fig:wAdjGraph2}.
Continuing to use the Hamming edit distance,
to achieve $K = 3$, $L = 3$, and the $M$ lower bound simultaneously
would require the embeddability of the graph of Fig.~\ref{fig:wAdjGraph2}
into the hypercube of size $3$.
Such embedding is clearly not possible since two nodes of the adjacency
graph have degree $4$, whereas the maximum degree of the hypercube is $3$.

To achieve the least increase in $M$ above the lower bound
(\ref{eqn:GraphEmbedLowerbound}), we must advantageously choose edges
in the adjacency graph to break to create embeddability.
(As we will see later, choosing the optimal set of edges to break involves
solving the error-tolerant subgraph isomorphism problem.)
In this example, the two nodes of degree $4$ must
each have at least one edge broken.  Picking the lowest weight edges
(the two with weight $1/10$) is clearly the best choice, as the resulting
graph can be embedded in the hypercube and cost of the edits
{\fransdvng k}$\leftrightarrow${\fransdvng G}
and
{\fransdvng c}$\leftrightarrow${\fransdvng J}
is increased by the least possible amount (from $1$ to $2$).
Each of the broken edges has probability $\frac{1}{10}\cdot\frac{1}{8}$,
so $M$ is increased above the previously computed minimum by $\frac{1}{40}$.
Thus we achieve $(K,L,M) = (3,3,\frac{19}{40})$.

\begin{figure}
  \centering
  \includegraphics[width=2.4in]{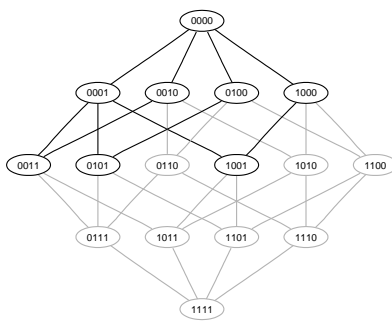}
  \caption{Weighted adjacency graph for editing process (\ref{eqn:editProcess2})
    embedded in 4-dimensional hypercube.  Black lines represent edges that are
    used in the embedding.  Gray lines represent edges in the hypercube that
    are unused in the embedding.}
  \label{fig:wAdjGraph2_hypercube}
\end{figure}

We may alternatively aim for lower $M$ at the expense of $K$ and $L$.
To determine whether the lower bound (\ref{eqn:GraphEmbedLowerbound})
can be achieved with $K = L = 4$, we need to check if the weighted adjacency
graph of Fig.~\ref{fig:wAdjGraph2} can be embedded in the hypercube
of size $4$.
Fig.~\ref{fig:wAdjGraph2_hypercube} shows that this embedding is possible,
with the code given in Table~\ref{table:code4hypercube}.
Thus one can achieve $(K,L,M) = (4,4,\frac{9}{20})$.

\begin{table}
 \caption{Code for the 4-dimensional hypercube embedding shown in
    Fig.~\ref{fig:wAdjGraph2_hypercube}.}
 \label{table:code4hypercube}
  \centering
    \begin{tabular}{|c|c|c|c|}
      \hline
{\fransdvng k} & $0000$  &
{\fransdvng K} & $0100$  \\ \hline
{\fransdvng G} & $1000$  &
{\fransdvng g} & $0010$  \\ \hline
{\fransdvng j} & $0011$  &
{\fransdvng J} & $1001$  \\ \hline
{\fransdvng C} & $0101$  &
{\fransdvng c} & $0001$  \\ \hline
    \end{tabular}
\end{table}

\subsection{Extension to Non-equiprobable Sources}
\label{sec:sourceGraphEmbed2}
\begin{figure}
  \centering
  \includegraphics[width=3in]{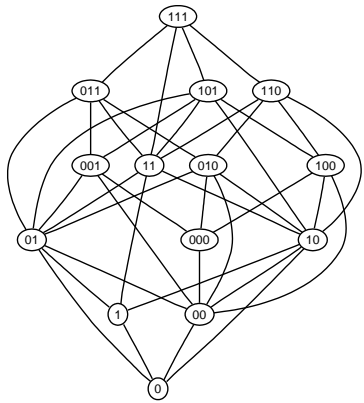}
  \caption{Levenshtein Edit Distance Graph for $\{0,1\}\cup\{0,1\}^2\cup\{0,1\}^3$.}
  \label{fig:LevenshteinGraph}
\end{figure}

The fact that both versions in the previous example were equiprobable and 
thus uncompressible might cast doubt on its gravity.  Here we consider 
another example where the sources are not equiprobable.  We will make use
of variable-length lossless source codes and the Levenshtein distance
as the edit distance.  The basic edit operations are substitution, insertion, 
and deletion, as opposed to the Hamming distance where substitution is the 
only edit operation.  Similar to the hypercube graph for the Hamming 
distance, we can create a Levenshtein edit distance graph.  The Levenshtein 
graph of binary strings up to length $3$ is shown in 
Fig.~\ref{fig:LevenshteinGraph}.  

Consider a memoryless source with alphabet $\mathcal{W} = \{${\fransdvng k},
{\fransdvng K}, {\fransdvng G}, {\fransdvng g}$\}$, 
with probabilities shown in Table~\ref{tab:Huff}.  Also in Table~\ref{tab:Huff}, 
we find a Huffman code for the source, which is the best variable-length 
lossless source code \cite{Huffman1952}.  
Since the marginal distribution $p(x)$ is dyadic, it is at the 
center of a code attraction region of the binary Huffman code
and achieves the entropy lower bound exactly \cite{LongoG1982}:
\[
K = \sum\limits_{x\in\mathcal{W}} p(x)\ell(x) = 1.75 = H(X) = -\sum\limits_{x\in\mathcal{W}} p(x)\log_2 p(x) \mbox{.}
\]

\begin{table}
 \caption{Huffman Code for $4$-ary source.}
 \label{tab:Huff}
  \centering
    \begin{tabular}{|c|c|c|c|c|}
      \hline
      $x \in \mathcal{W}$& $p(x)$& $f_{\mbox{Huffman}}(x)$ & $\ell_{\mbox{Huffman}}(x)$& $p(x)\ell(x)$ \vspace{.5mm}\\ \hline
      {\fransdvng k} & $\tfrac{1}{2}$ & $0$ & $1$ & $\tfrac{1}{2}$ \\ \hline
      {\fransdvng K} & $\tfrac{1}{4}$ & $10$ & $2$ & $\tfrac{1}{2}$ \\ \hline
      {\fransdvng G} & $\tfrac{1}{8}$ & $110$ & $3$ & $\tfrac{3}{8}$ \\ \hline
      {\fransdvng g} & $\tfrac{1}{8}$ & $111$ & $3$ & $\tfrac{3}{8}$ \\ \hline
    \end{tabular}
\end{table}

\begin{figure}
  \centering
  \includegraphics[width=1.5in]{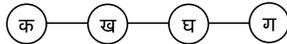}
  \caption{Adjacency graph for noisy typewriter-like channel.}
  \label{fig:wAdjGraphHuffman}
\end{figure}

\begin{figure}
  \centering
  \includegraphics[width=3in]{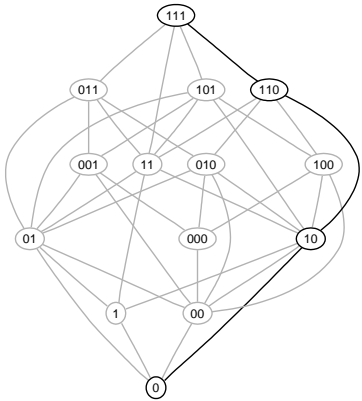}
  \caption{Adjacency graph for noisy typewriter-like channel embedded in the Levenshtein graph.}
  \label{fig:LevenshteinHuffmanGraph}
\end{figure}

Now consider a channel that is like the noisy typewriter channel,  
with channel transmission matrix 
\begin{equation}
p(y|x) = \left[ {\begin{array}{*{20}c}
   \tfrac{3}{4} & \tfrac{1}{2} & 0 & 0 \vspace{.5mm}\\
   \tfrac{1}{4} & \tfrac{3}{8} & \tfrac{1}{4} & 0 \vspace{.5mm}\\
   0 & \tfrac{1}{8} & \tfrac{1}{2} & \tfrac{1}{4} \vspace{.5mm}\\
   0 & 0 & \tfrac{1}{4} & \tfrac{3}{4} \vspace{.5mm}\\
\end{array}} \right].
\end{equation}
Evidently the editing is stationary, so the same Huffman code is optimal for 
both $X$ and $Y$.  Constructing the adjacency graph yields 
Fig.~\ref{fig:wAdjGraphHuffman}.
This graph can be embedded (with matched vertex labels)
in the Levenshtein graph 
using the Huffman assignment that we had developed, 
as shown in Fig.~\ref{fig:LevenshteinHuffmanGraph}.  

Evaluating the malleability lower bound (\ref{eqn:GraphEmbedLowerbound})
for $n=1$ in this case gives
\[
M \ge \sum\limits_{x\in\mathcal{W}}\sum\limits_{y\in\mathcal{W}:y\neq x}p(x,y) = \tfrac{3}{8} \mbox{.}
\]
With the code that we have used, we can achieve the triple 
$(K,L,M) = \left(\tfrac{7}{4},\tfrac{7}{4},\tfrac{3}{8}\right)$
which meets the $n=1$ lower bounds tightly, so it is optimal in the compression 
and malleability senses. 
As before, we can consider Cartesian products to reduce $M$,
however, things are a bit more complicated since the Levenshtein graph does 
not grow as a Cartesian product. 

\subsection{Minimal Change Codes}
\label{sec:minimalchangecodes}
As seen in the previous subsection, Gray codes and related minimal change 
code constructions seem to play a role in achieving good palimpsest
performance.  We review minimal change codes and some of their 
previous uses in communication theory, pointing out connections to our problem.
We use minimal change codes to expand our treatment in the previous parts
from using just Hamming or Levenshtein distances to include general edit distances.

\begin{defn}
Let $G$ be a connected graph.  The \emph{path metric} $d_G$ associated with the
graph $G$ is the integer-valued metric on the vertices of $G$ which is defined
by setting $d_G(u,v)$ equal to the length of the shortest path in $G$ joining
$u$ and $v$.
\end{defn}
\begin{prop}
For any edit distance $d: \mathcal{V}^{*}\times\mathcal{V}^{*} \to [0,\infty)$, 
there exists a graph $G$ with vertex set $\mathcal{V}^{*}$ such that its path metric
$d_G = d$.
\end{prop}
\begin{IEEEproof}
Construct a graph on vertex set $\mathcal{V}^{*}$ by adding an edge for any pair 
of vertices $A,B \in \mathcal{V}^{*}$ such that $d(A,B) = 1$.  
\end{IEEEproof}
\begin{defn}
An ordered codebook $(A_i)$, $i=1,2,\ldots,|(A_i)|$,
$A_i \in \mathcal{V}^{*}$ is
a \emph{minimal change code} with respect to edit distance $d$
  if it is a Hamiltonian path in a subgraph of the graph on
  $\mathcal{V}^{*}$ associated with $d$.
\end{defn}

Our definition of minimal change codes is a generalization of Gray codes, which
are Hamiltonian paths through the hypercube associated with Hamming distance 
\cite{Gilbert1958}.  Other minimal change codes may include Hamiltonian paths 
through the Levenshtein graph (Fig.~\ref{fig:LevenshteinGraph}), the de Bruijn 
graph, or the graph induced by Dobrushin's distance functions for 
insertion/deletion channels \cite{Dobrushin1967}.  There are countless
other edit distances with numerous minimal change codes corresponding to each.

Minimal change codes have been used previously in the architecture design of
parallel computers and in switching theory, among other places.  Of particular 
interest to us, however, is their use in joint source-channel coding (JSCC)
\cite{ZegerG1990}.  There are related problems in signal constellation labeling 
\cite{AgrellLSO2004,CaireTB1998,SethuramanH2006}, in the genotype to phenotype 
mapping problem mentioned previously \cite{Swanson1984,Tlusty2007}, or in the problem of 
labeling books for ease of browsing \cite{Losee1992}.  There are also several
theories of cognition based on preserving similarity relations from a source space
in a representation space, though minimal change codes do not seem to be used 
explicitly \cite{Edelman1999}.

Consider JSCC with source alphabet $\mathcal{X}$, channel input alphabet
$\mathcal{A}$, channel output alphabet $\mathcal{B}$, and source
reconstruction alphabet $\mathcal{Y}$.  Then the injective mapping
between $\mathcal{X}$ and $\mathcal{A}$ is the index assignment for
JSCC.  The mapping between $\mathcal{A}$ and 
$\mathcal{B}$ is given by the noisy channel, a transition probability 
assignment, $p(b|a)$.  The surjective mapping between $\mathcal{B}$ 
and $\mathcal{Y}$ is the inverse index assignment operation.  The goal 
in selecting index assignments is to minimize the distortion between the
$\mathcal{X}-\mathcal{Y}$ spaces when there are errors between the
$\mathcal{A}-\mathcal{B}$ spaces.  Informally using terminology from
genetics, the source spaces $\mathcal{X}$ and $\mathcal{Y}$ are cast 
as phenotype, whereas the index spaces $\mathcal{A}$ and
$\mathcal{B}$ are cast as genotype.  Then index assignment aims to
have small mutations in genotype result in small changes in phenotype.

In the palimpsest problem,\footnote{To make the correspondence 
more precise, let $\mathcal{X} = \mathcal{Y} = \mathcal{W}$ and 
$\mathcal{A} = \mathcal{B} = \mathcal{V}$.} the injective and surjective 
mappings between $\mathcal{X}$ and $\mathcal{A}$ as well as $\mathcal{B}$ 
and $\mathcal{Y}$ are basically the same as in the joint source channel 
coding problem.  The distinction between the two problems is that for 
malleable coding, there is a transition probability assignment 
between $\mathcal{X}$ and $\mathcal{Y}$, rather than between $\mathcal{A}$ 
and $\mathcal{B}$.  One goal is to minimize the distance between words 
in the $\mathcal{A}$ and $\mathcal{B}$ spaces for perturbations in 
the $\mathcal{X}$ and $\mathcal{Y}$ spaces.  Using the genetics 
analogy, index assignments so that small changes in phenotype result in small 
changes in genotype are desired.  One might even call malleable coding
a problem in joint channel source coding.

Considering that index assignment for JSCC, 
signal constellation labeling, and the palimpsest problem are so 
similar, it is not surprising that Gray codes come up in all of them
\cite{ZegerG1990,AgrellLSO2004}.  All are essentially problems of 
embedding: performing a transformation on objects of one type to
produce objects of a new type such that the distance between the
transformed objects approximates the distance between the original
objects~\cite{Cormode2003}.  

\section{General Characterizations}
\label{sec:palimpSol}
We have seen that there may be a trade-off between the various 
parameters $(K,L,M)$ and have found several easily achieved points.
Our interest now turns to obtaining more detailed characterizations
of $\mathfrak{P}_V$ and $\mathfrak{P}_B$, the sets of achievable
rate--malleability triples.

\subsection{Variable-length Coding}
\label{sec:var_ach}
We begin with characterization of $\mathfrak{P}_V$,
which is a problem in zero error information theory \cite{AlonO1996,KornerO1998}.
Our results are expressed in terms of the solution to an error-tolerant 
attributed subgraph isomorphism problem \cite{MessmerB1998},
which we first describe in general.  

\subsubsection{Error-Tolerant Attributed Subgraph Isomorphism Problem}
A vertex-attributed graph is a three-tuple $G = (V,E,\mu)$, where
$V$ is the set of vertices, $E \subseteq V\times V$ is the set 
of edges, and $\mu:V\to \mathcal{V}^{*}$ is a function assigning labels to
vertices.  The set of labels is denoted $\mathcal{V}^{*}$.  The definition of
embedding for attributed graphs has a slightly stronger requirement
than for unattributed graphs, Def.~\ref{def:embeddable}.  

\begin{defn}
\label{def:a_embeddable}
Consider two vertex-attributed graphs $G = (V(G),E(G),\mu_G)$ and 
$H = (V(H),E(H),\mu_H)$.  Then $G$ is said to be \emph{embeddable}
into $H$ if $H$ has a subgraph isomorphic to $G$.  That is, there is an 
injective map $\phi: V(G) \to V(H)$ such that $\mu_G(v) = \mu_H(\phi(v))$
for all $v \in V(G)$ and that $(u,v) \in E(G)$ implies 
$(\phi(u),\phi(v)) \in E(H)$.  This is denoted as $G \leadsto H$.
\end{defn}

Several graph editing operations may be defined, such as substituting
a vertex label, deleting a vertex, deleting an edge, and inserting an edge.
These four operations are powerful enough to transform any attributed 
graph into a subgraph of any other attributed graph.  The edited graph
is denoted through the operator $\mathcal{E}(\cdot)$ corresponding to the
sequence of graph edit operations $\mathcal{E} = (e_1,\ldots,e_k)$.  
There is a cost associated with each sequence of graph edit operations,
$C(\mathcal{E})$.  

\begin{defn}
Given two graphs $G$ and $H$, an \emph{error-correcting attributed
subgraph isomorphism} $\psi$ from $G$ to $H$ is the composition of
two operations $\psi = (\mathcal{E},\phi_{\mathcal{E}})$ where
\begin{itemize}
  \item $\mathcal{E}$ is a sequence of graph edit operations such that there exists a $\mathcal{E}(G)$ that satisfies $\mathcal{E}(G) \leadsto H$.
  \item $\phi_{\mathcal{E}}$ is an embedding of $\mathcal{E}(G)$ into $H$.
\end{itemize}
\end{defn}

\begin{defn}
The \emph{subgraph distance} $\rho(G,H)$ is the cost of the minimum cost 
error-correcting attributed subgraph isomorphism $\psi$ from $G$ to $H$.
\end{defn}
Note that in general, $\rho(G,H) \neq \rho(H,G)$.

\begin{rem}
It should be noted that the subgraph isomorphism problem is NP-complete \cite{GareyJ1979},
and therefore the error-tolerant subgraph isomorphism problem is in the class NP
and is generally harder than the exact subgraph isomorphism problem \cite{MessmerB1998}.
\end{rem}

\subsubsection{Closeness Vitality}
The subgraph isomorphism cost structure for the palimpsest problem 
is based on a graph theoretic quantity called the 
closeness vitality \cite{KoschutzkiLPRTZ2005}.
Vitality measures determine the importance of particular edges and vertices
in a graph.  
\begin{defn}
Let $\mathfrak{G}$ be the set of all graphs $G = (V,E)$,
and let $f: \mathfrak{G}\to\mathbb{R}$
be any real-valued function on $\mathfrak{G}$.  A vitality index $v(G,x)$ is the 
difference of the values of $f$ on $G$ and on $G$ without element $x$; it satisfies $v(G,x) = f(G) - f(G-x)$.
\end{defn}
A particular vitality index is the closeness vitality, defined in terms
of the Wiener index \cite{Wiener1947}, which is simply the sum of all pairwise 
distances.  
\begin{defn}
The \emph{Wiener index} $f_W(G)$ of a graph $G$ is the sum of the distances 
of all vertex pairs:
\[
f_W(G) = \sum_{v\in V}\sum_{w\in V} d(v,w) \mbox{.}
\]
\end{defn}
\begin{defn}
The \emph{closeness vitality} $\cv(G,x)$ is the vitality index with respect to the Wiener index:
\[
\cv(G,x) = f_W(G) - f_W(G-x) \mbox{.}
\]
\end{defn}
In addition to the application in the palimpsest problem, the closeness 
vitality also determines traffic-related costs in all-to-all routing networks.

Finding the distance matrix to compute the Wiener index 
involves solving the all-pairs shortest path problem.  Finding 
the distance matrix of a modified graph from the distance matrix of 
the original graph involves solving the dynamic all-pairs shortest path problem 
\cite{AusielloISN1991,DemetrescuI2004}.  

\subsubsection{$\mathfrak{P}_V$ Characterization}
For our purposes, we are concerned with the error-tolerant embedding of
an attributed, weighted source adjacency graph into the graph induced by 
a $\mathcal{V}^{*}$-space edit distance.  As such, edge deletion will be 
the only graph editing operation that is required.  Error-tolerant
embedding problems in pattern recognition and machine vision often have simple
cost functions \cite{MessmerB1998,Bunke1999}; our cost function is determined 
by the closeness vitality and is not so simple.

To characterize $\mathfrak{P}_V$, 
let us first consider the delay-free case, $n=1$.  A source $p(X,Y)$
and an edit distance $d(\cdot,\cdot)$ are given.
It is known \cite{Huffman1952} that 
Huffman coding provides the minimal redundancy instantaneous code and achieves
expected performance $H(X) \le K \le H(X) + 1$.  
Similarly, a Huffman code for $Y$ would yield $H(Y) \le L \le H(Y) + 1$. 
The rate loss for using an incorrect Huffman code is essentially
as given in Fig.~\ref{fig:differentmarginals}.  Suppose that we 
require that the rate lower bound is met, i.e. we must use a Huffman 
code for some $Z$ that is on the geodesic between $X$ and $Y$.
This code will satisfy the Kraft inequality \cite{Kraft1949}.
Note that for a given $Z$, there are several Huffman codes:
those arising from different labelings of the code tree and
also perhaps different trees \cite{Ahlswede2006}.  Let us
denote the set of all Huffman codes for $Z$ as $\mathcal{H}_Z$.

Since $K$ and $L$ are fixed by the choice of $Z$, all that remains is
to determine the set of achievable $M$.  Let $G$ be the graph induced
by the edit distance $d(\cdot,\cdot)$, and $d_G$ its path metric.  
The graph $G$ is intrinsically labeled. 
Let $A$ be the weighted adjacency graph of the source $p(X,Y)$, with
vertices $\mathcal{W}$, edges $E(A)$ a subset of 
$\mathcal{W}\times\mathcal{W}$, and labels given by a Huffman code.
That is $A = (\mathcal{W},E(A),f_E)$ for some $f_E\in\mathcal{H}_Z$.
There is a path semimetric, $d_A$, associated with the graph $A$ 
(since the adjacency graph is weighted, it might not satisfy the triangle inequality).

As may be surmised from Section~\ref{sec:ConstPalimEx}, the basic problem is to 
solve the error-tolerant subgraph isomorphism problem of embedding $A$ into $G$.
In general for $n=1$, the malleability cost under edit distance $d_G$ 
when using the source code $f_E$ is
\[
M = \sum_{x\in\mathcal{W}}\sum_{y\in\mathcal{W}}p(x,y) d_G(f_E(x),f_E(y)) \mbox{.}
\]
The smallest malleability possible is when $A = (\mathcal{W},E(A),f_E)$ is
a subgraph of $G$, and then 
\[
M_{min} = \sum_{x\in\mathcal{W}}\sum_{y\in\mathcal{W}}p(x,y) d_A(x,y) = \sum_{x\in\mathcal{W}}\sum_{y\in\mathcal{W}}p(x,y) d_G(f_E(x),f_E(y)) = \Pr[X \neq Y] \mbox{,}
\]
which is simply the expected Wiener index
\[
M_{min} = E[ f_W(A)] = \Pr[X \neq Y] \mbox{.}
\]

If edges in $A$ need to broken in order to make it a subgraph of $G$, 
then $M$ increases as a result.
The cost of graph editing operations in the error-tolerant 
embedding problem should reflect the effect on $M$.  If an edge $\bar{e}$ is 
removed from the graph $A$, the resulting graph is called $A-\bar{e}$; it 
induces its own path semimetric $d_{A-\bar{e}}$.  Thus the cost of removing an 
edge, $\bar{e}$, from the graph $A$ is given by the following expression
as a function of the associated removal operation $e$:
\[
C(e) = \sum_{x\in\mathcal{W}}\sum_{y\in\mathcal{W}}p(x,y) \left[ d_{A-\bar{e}}(f_E(x),f_E(y)) - d_A(f_E(x),f_E(y)) \right]\mbox{,}   
\]
which is the negative expected closeness vitality
\[
C(\delta) = -E [\cv(A,e)].
\]
If $\mathcal{E}$ is a sequence of edge removals, $\bar{\mathcal{E}}$, then
\[
C(\mathcal{E}) = \sum_{x\in\mathcal{W}}\sum_{y\in\mathcal{W}}p(x,y) \left[d_{A-\bar{\mathcal{E}}}(f_E(x),f_E(y)) - d_{A}(f_E(x),f_E(y))\right] \mbox{,}  
\]
which is
\[
C(\mathcal{E}) = -E [\cv(A,\bar{\mathcal{E}})] \mbox{.}
\]

As seen, the cost function is quite different from standard error-tolerant
embedding problems \cite{MessmerB1998,Bunke1999} since it depends 
not only on which edge is broken, but also on the remainder of 
the graph.

Putting things together, we see that $\mathfrak{P}_V$ contains any point
\[
(K,L,M) = \left(H(X) + D(p_X\|p_Z) + 1, H(Y) + D(p_Y\|p_Z) + 1, M_{min} + \mathop{\min}\limits_{f_E\in\mathcal{H}_Z} \rho(A,G)  \right) \mbox{.}
\]
The previous analysis had assumed $n = 1$.  We may increase the block length
and improve performance.  

\begin{thm}
\label{thm:subgraphIsoAchievable}
Consider a source $p(X,Y)$ with associated (unlabeled) weighted adjacency graph $A$ and 
an edit distance $d$ with associated graph $G$.  
For any $n$, let $\mathfrak{P}_V^{(ach)}$ be the set of triples $(K,L,M)$ that are computed, 
by allowing an arbitrary choice of the memoryless random variable $p(Z_1^n)$, 
as follows: 
\[
K = H(X) + D(p_X\|p_Z) + \tfrac{1}{n}\mbox{,}
\]
\[
L = H(Y) + D(p_Y\|p_Z) + \tfrac{1}{n}\mbox{,}
\]
\[
M = \tfrac{1}{n} \Pr[X_1^n \neq Y_1^n] + \tfrac{1}{n}\mathop{\min}\limits_{f_E\in\mathcal{H}_{Z_1^n}} \rho(A = (\mathcal{W}^n,E(A),f_E),G) \mbox{.}
\]
Then the set of triples 
$\mathfrak{P}_V^{(ach)} \subseteq \mathfrak{P}_V$ is achievable instantaneously.  
\end{thm}
\begin{IEEEproof}
A non-degenerate random variable $Z_1^n$ is fixed.  
There is a family of instantaneous lossless codes (with $\Delta = 0$) that corresponds 
to this random variable, denoted $\{(f_E,f_D)\} = \mathcal{H}_{Z_1^n}$, through the McMillan sum.  
By the results in \cite{Gilbert1971}, any of these codes achieve rates $K \le H(X) + D(p_X\|p_Z) + \tfrac{1}{n}$
and $L \le H(Y) + D(p_Y\|p_Z) + \tfrac{1}{n}$.  Moreover, by the graph embedding construction,
a code $(f_E,f_D)$ achieves $M = \tfrac{1}{n} \Pr[X_1^n \neq Y_1^n] + \tfrac{1}{n} \rho(A = (\mathcal{W}^n,E(A),f_E),G)$.
Since all codes in $\mathcal{H}_{Z_1^n}$ have the same rate performance, a code in the family
that minimizes $\rho$ may be chosen.
\end{IEEEproof}
The above theorem states that error-tolerant subgraph isomorphism implies achievable malleability.
The choice of the auxiliary random variable $Z$ is open to optimization.  If minimal rates
are desired, then $p_Z$ should be on the geodesic connecting $p_X$ and $p_Y$.  If $Z$ is not
on the geodesic, then there is some rate loss, but perhaps there can be some malleability gains.

Note that when $p(y|x)$ is a stationary editing process, there is the possibility
of the simple lower bounds being tight to this achievable region.
\begin{cor}
Consider a source as given above in Theorem \ref{thm:subgraphIsoAchievable}.
If $p(y|x)$ is stationary, $p(x) = p(y)$ is $|\mathcal{V}|$-adic, and there is a 
Huffman-labeled $A$ for $p(x) = p(y)$ that is an isometric subgraph of $G$, then the
block length $n$ lower bound $(H(X),H(Y),\tfrac{1}{n}\Pr[X_1^n \neq Y_1^n])$ is tight to this 
achievable region for every $n$, and in particular to $(H(X),H(Y),0)$ for large $n$.  
\end{cor}

\subsection{Block Coding}
\label{sec:block_ach}
Now we turn our attention to the block-coding palimpsest problem.  
For $\mathfrak{P}_B$, we use a joint typicality graph rather than
the weighted adjacency graph used for $\mathfrak{P}_V$.  Additionally 
we focus on binary block codes under Hamming edit distance, so we are 
concerned only with hypercubes rather than general edit distance graphs.  

We can use graph-theoretic ideas to formally state an achievability
result for the block coding palimpsest problem.  
As shown in the constructive examples above, there are schemes for
which an improvement on $M$ may be achieved by increasing $L$.  
However, the resulting compression of $Y_1^n$ is not unique, 
and thus is not optimal.  We wish to expurgate the redundant representations of
$Y_1^n$ as efficiently as possible, by the aid of a graph.
However, in doing so, we have to also consider the representations
and how they are related to one another.  First we review some standard
typicality arguments (from \cite{Yeung2002}) and then define a graph 
from typical sets.

\begin{defn}
The strongly typical set $T^n_{[X]\delta}$ with respect to $p(x)$ is
\[
T^n_{[X]\delta} = \left\{x_1^n \in \mathcal{X}^n \mid \sum_x \left| \tfrac{1}{n} N(x;x_1^n) - p(x)\right| \le \delta \right\} \mbox{,}
\]
where $N(x;x_1^n)$ is the number of occurrences of $x$ in $x_1^n$ and $\delta > 0$.
\end{defn}

\begin{defn}
The strongly jointly typical set $T^n_{[XY]\delta}$ with respect to $p(x,y)$ is
\[
T^n_{[XY]\delta} = \left\{(x_1^n,y_1^n) \in \mathcal{X}^n \times \mathcal{Y}^n \mid \sum_x \sum_y \left| \tfrac{1}{n} N(x,y;x_1^n,y_1^n) - p(x,y)\right| \le \delta \right\} \mbox{.}
\]
\end{defn}

\begin{defn}
For any $x_1^n \in T^n_{[X]\delta}$, define a strongly conditionally typical set
\[
T^n_{[Y|X]\delta}(x_1^n) = \left\{y_1^n \in T^n_{[Y]\delta} \mid (x_1^n,y_1^n) \in T^n_{[XY]\delta} \right\} \mbox{.}
\]
\end{defn}

\begin{defn}
Let the connected strongly typical set be
\[
S^n_{[X]\delta} = \left\{x_1^n \in T^n_{[X]\delta} \mid T^n_{[Y|X]\delta}(x_1^n) \mbox{ is nonempty}\right\} \mbox{.}
\]
\end{defn}

Now that we have definitions of typical sets, we put forth some lemmas.
\begin{lem}[Strong AEP]
\label{lemma:StrongAEP}
Let $\eta$ be a small positive number such that $\eta \to 0$ as $\delta \to 0$.  Then
for sufficiently large $n$,
\[
\left|T^n_{[X]\delta}\right| \le 2^{n(H(X)+\eta)} \mbox{.}
\]
\end{lem}
\begin{IEEEproof}
See \cite[Theorem 5.2]{Yeung2002}.
\end{IEEEproof}

\begin{lem}[Strong JAEP]
\label{lem:jaep}
Let $\lambda$ be a small positive number such that $\lambda \to 0$ as $\delta \to 0$.  Then
for sufficiently large $n$,
\[
\Pr[(X_1^n,Y_1^n) \in T^n_{[XY]\delta}] > 1 - \delta
\]
and
\[
(1-\delta)2^{n(H(X,Y)-\lambda)} \le \left|T^n_{[XY]\delta}\right| \le 2^{n(H(X,Y)+\lambda)} \mbox{.}
\]
\end{lem}
\begin{IEEEproof}
See \cite[Theorem 5.8]{Yeung2002}.
\end{IEEEproof}
\begin{lem}
\label{lem:jaepval}
If $\delta(n)$ satisfies the following conditions, then Lemma \ref{lem:jaep} remains valid:
\[
\delta(n) \to 0 \mbox{ and } \sqrt{n}\delta(n) \to \infty \mbox{ as } n\to\infty \mbox{.}
\]
\end{lem}
\begin{IEEEproof}
See \cite[(2.9) on p. 34]{CsiszarK1997}.
\end{IEEEproof}
\begin{lem}
\label{lem:caep}
If $\left|T^n_{[Y|X]\delta}(x_1^n)\right| \ge 1$, then
\[
2^{n(H(Y|X)-\nu)} \le \left|T^n_{[Y|X]\delta}(x_1^n)\right| \le 2^{n(H(Y|X)+\nu)} \mbox{,}
\]
where $\nu \to 0$ as $n \to \infty$ and $\delta \to 0$.
\end{lem}
\begin{IEEEproof}
See \cite[Theorem 5.9]{Yeung2002}.
\end{IEEEproof}

\begin{lem}
\[
(1-\delta)2^{n(H(X)-\psi)} \le \left|S^n_{[X]\delta}\right| \le 2^{n(H(X)+\psi)} \mbox{,}
\]
where $\psi \to 0$ as $n \to \infty$ and $\delta \to 0$.  Also, for any $\delta > 0$, 
\[
\Pr[X_1^n \in S^n_{[X]\delta}] > 1 - \delta
\]
for $n$ sufficiently large.
\end{lem}
\begin{IEEEproof}
See \cite[Corollary 5.11]{Yeung2002}, Lemma \ref{lemma:StrongAEP}, and \cite[Proposition 5.12]{Yeung2002}.
\end{IEEEproof}

For the bivariate 
distribution $p(x,y)$, define a square matrix called the 
\emph{strong joint typicality matrix} $A^n_{[XY]}$ as follows.
There is one row (and column) for each sequence in
$S^n_{[X]\delta} \cup S^n_{[Y]\delta}$.  
The entry with row corresponding to $x_1^n$ and column corresponding to $y_1^n$
receives a one if $(x_1^n,y_1^n)$ is strongly jointly 
typical and zero otherwise.

\subsubsection{Stationary Editing}
Now let us restrict ourselves to the class of bivariate distributions
with equal marginals:
\[
\mathcal{P} = \left\{p(x,y) \mid p(x) = p(y)\right\}\mbox{,}
\]
which is the class of distributions with stationary editing.  
In this class, we avoid the mismatch redundancy and also reduce the 
number of performance parameters from 3 to 2.  After this restriction, 
it is clear that the $x$-typical set and the $y$-typical set coincide.  
Moreover, $H(X)=H(Y)$ and $H(Y|X) = H(X|Y) = H(X)-I(X;Y)$.  Thus it 
follows that asymptotically, $A^n_{[XY]}$ will be a square matrix with 
an approximately equal number of ones in all columns and in all rows.  Think of $A^n_{[XY]}$
as the adjacency matrix of a graph, where the vertices are sequences 
and edges connect sequences that are jointly typical with one another.  
\begin{prop}
Take $A^n_{[XY]}$ for some source in $\mathcal{P}$ as the adjacency matrix of 
a graph $\mathcal{G}^n$.  The number of vertices in the graph will satisfy
\[
(1-\delta)2^{n(H(X)-\psi)} \le \left|V(\mathcal{G}^n)\right| \le 2^{n(H(X)+\psi)} \mbox{,}
\]
where $\psi \to 0$ as $n \to \infty$ and $\delta \to 0$.  The degree of each
vertex, $\mbox{deg}_v$, will concentrate as
\[
2^{n(H(Y|X)-\nu)} \le \mbox{deg}_v \le 2^{n(H(Y|X)+\nu)} \mbox{,}
\]
where $\nu \to 0$ as $n\to\infty$ and $\delta\to 0$.
\end{prop}
\begin{IEEEproof}
Follows from the previous lemmas.
\end{IEEEproof}

Having established the basic topology of the strongly typical set as asymptotically a 
$2^{nH(Y|X)}$-regular graph on $2^{nH(X)}$ vertices, we return to the 
coding problem.  Using graph embedding ideas yields a theorem on
block palimpsest achievability.
\begin{thm}
\label{thm:achBlockPalimpsest}
For a source $p(x,y) \in \mathcal{P}$ and the Hamming edit distance,
a triple $(K,K,M = M_{min})$ is achievable if $\mathcal{G}^n \leadsto H_{nK}$.
\end{thm}
\begin{IEEEproof}
To achieve $M_{min}$, we need to assign binary codewords to each of 
the $2^{nH(X)}$ vertices, such that the Hamming distance between the 
codeword of a vertex and the codewords of any of its neighbors is $1$.  
Using the binary reflected Gray code of length $nK$ and the hypercube that 
it induces, the construction reduces
to finding an embedding of $\mathcal{G}^n$ into the hypercube of size $nK$,
denoted $H_{nK}$.  Thus a sufficient condition for block-code 
achievability, while requiring $M = M_{min}$, 
is $\mathcal{G}^n \leadsto H_{nK}$.
\end{IEEEproof}

Using this result, we argue that a linear increase in malleability is at
exponential cost in code length.  By a simple counting argument, we present 
a condition for embeddability. 
\begin{thm}
\label{thm:expBlockPalimpsest}
For a source $p(x,y) \in \mathcal{P}$ and the Hamming edit distance, 
asymptotically, if $\mathcal{G}^n \leadsto H_{nK}$ then
\begin{equation}
nK \ge \max\left(nH(X),2^{nH(Y|X)}\right) \mbox{.}
\end{equation}
\end{thm}
\begin{IEEEproof}
The hypercube $H_{nK}$ is an $nK$-regular graph with $2^{nK}$ vertices.  As a 
minimal condition for embeddability, the number of vertices in the 
hypercube must be greater than or equal to the number of vertices in 
the graph to be embedded, i.e. $nK \ge n(H(X)+\psi)$.  As another minimal 
condition for embeddability, the degree of the hypercube must be greater 
than or equal to the maximal degree of the graph to be embedded, so
$nK \ge 2^{n(H(Y|X)+\nu)}$.  Combining the two conditions and letting
$\psi \to 0$ and $\nu \to 0$ as $n\to \infty$ yields the desired result.
\end{IEEEproof}

This theorem is one of our main results.
It should be noted that even if we allowed some asymptotically small slack in breaking some 
edges to perform embedding, i.e.\ we solved an error-tolerant subgraph 
isomorphism problem with error tolerance $\xi$, this would not help,
since we would need to break a constant fraction of edges in $\mathcal{G}^n$ 
to reduce the maximal degree.  In particular, since each of the $nH(X)$ vertices in $\mathcal{G}^n$ 
asymptotically has the same degree, to reduce the maximal degree even by one would require breaking 
$\xi \ge nH(X)$ edges.  Clearly $\xi \nrightarrow 0$ as $n\to\infty$. 

This result can be interpreted as follows.  When using binary codes that 
achieve the minimal malleability parameter, the length of the code must be 
greater than $\max\left(nH(X),2^{nH(Y|X)}\right)$.  If $2^{nH(Y|X)}$ is 
much greater than $nH(X)$, i.e. the two versions are not particularly 
well correlated, this implies that to achieve minimal malleability 
requires a significant length expansion of the codewords over the entropy 
bound.  Taking this to an extreme, suppose that $X$ and $Y$ are independent.  
Then $2^{nH(Y|X)} = 2^{nH(X)}$, and an exponential expansion is required,
just as in the universal PPM scheme of Section~\ref{sec:PPM}.  

If we want to understand the embeddability requirements further, we would 
need to understand the topology of $\mathcal{G}^n$ further.  Just knowing 
that it is asymptotically regular does not seem to be enough.  Several
properties that are equivalent to exact hypercube embeddability are given in 
\cite{DezaL1997,DezaGS2004}.\footnote{There are several 
characterizations of hypercube-embeddable graphs in the 
metric theory of graphs \cite{DezaL1997,DezaGS2004}.
For a bipartite connected graph $G = (V,E)$ the statements are equivalent:
\begin{itemize}
  \item $G$ can be isometrically embedded into a hypercube.
  \item $G$ satisfies $G(a,b) = \{x\in V | d_G(x,a) < d_G(x,b)\}$ is 
	convex for each edge $(a,b)$ of $G$.  A subset $U \subseteq V$ is convex 
	if it is closed under taking shortest paths.
  \item $G$ is an $\ell_1$ graph, i.e.\ the path metric $d_G$ is isometrically embeddable in the space $\ell_1$.
  \item The path metric $d_G$ satisfies the pentagonal inequality:
	\[
	d_G(v_1,v_2) + d_G(v_1,v_3) + d_G(v_2,v_3) + d_G(v_4,v_5) - \sum\limits_{h=1,2,3 \mbox{  }k=4,5} d_G(v_h,v_k) \le 0
	\]
	for all nodes $v_1,\ldots,v_5 \in V$.
  \item The distance matrix of $G$ has exactly one positive eigenvalue.
\end{itemize}
Further, a graph is said to be distance regular 
if there exist integers $b_m,c_m$ ($m>0$) such that for any two nodes 
$i,j \in V(G)$ at distance $d_G(i,j) = m$ there are exactly $c_m$ nodes at 
distance $1$ from $i$ and distance $m-1$ from $j$, and there are $b_m$ nodes 
at distance $1$ from $i$ and distance $m+1$ from $j$.  The distance-regular 
graphs that are hypercube embeddable are completely classified: 
the hypercubes, the even circuits, and the double-odd graphs.}  Of course
we can break some small fraction $\xi$ of edges in the graph to satisfy
the embeddability conditions as long as $\xi \to 0$ as $n\to\infty$.
If we no longer require that $nM$ be the minimal possible, then we
are back to the same kind of error-tolerant subgraph isomorphism 
formulation given for variable length coding in the previous section.
The only change in the characterization of the achievable region
is that rather than restricting the encoder to be the Huffman code
of an auxiliary random variable $Z_1^n$, here one would need to test
the error-tolerant subgraph isomorphism functional over all permutations
of labelings.

\subsubsection{General Editing}
If we remove our restriction of $p(x,y) \in \mathcal{P}$, then we can
create $A^n_{[XY]}$ as before.
While the 
resulting graph would not be asymptotically regular, the basic result
on paying an exponential rate penalty will still hold.

The space $S^n \triangleq S^n_{[X]\delta} \cup S^n_{[Y]\delta}$ 
with the corresponding path metric, $d_A$ induced by $A^n_{[XY]}$ is a metric space.
Hypercubes with their natural path metric, $d_G$, are also a metric space.  Rather than
requiring absolutely minimal $nM$, it can be noted that $M$ is asymptotically zero
when the Lipschitz constant associated with the mapping between the source space
and the representation space has nice properties in $n$.  

\begin{defn}
A mapping $f$ from the metric space $(S^n,d_{A^n})$ to the metric
space $(\mathcal{V}^{nK},d_{G^n})$ is called \emph{Lipschitz continuous} if
\[
d_{G^n}(f(x_1),f(x_2)) \le C d_{A^n}(x_1,x_2)
\]
for some constant $C$ and for all $x_1,x_2 \in S^n$. 
The smallest such $C$ is the \emph{Lipschitz constant}:
\[
\Lip[f] = \sup\limits_{x_1 \neq x_2\in S^n} \frac{d_{G^n}\left(f(x_1),f(x_2)\right)}{d_{A^n}(x_1,x_2)} \mbox{.}
\]
\end{defn}

The Lipschitz constant is also called the \emph{dilation} of an embedding, since
it is the maximum amount that any edge in $\mathcal{G}^n$ is stretched as it is
replaced by a path in $H_{nK}$ \cite{RosenbergH2001,LivingstonS1988}.  A related
quantity is the Lipschitz constant of the inverse mapping, called the \emph{contraction}:
\[
\Lip[f^{-1}] = \sup\limits_{x_1 \neq x_2\in S^n} \frac{d_{A^n}(x_1,x_2)}{d_{G^n}\left(f(x_1),f(x_2)\right)} \mbox{.}
\]
The product of the dilation and contraction $\Lip[f]\Lip[f^{-1}]$ is called the \emph{distortion}.
Another property of metric embeddings in the \emph{expansion},
which is the ratio of the sizes of the two finite metric spaces, 
\[
\expan[f] = \frac{|\mathcal{V}^{nK}|}{|S^n|}\mbox{.}
\]

We can bound the malleability $M$, for a coding scheme that only represents 
sequences in $S^n$ as follows.
\begin{thm}
\label{thm:palimpsestLipsz}
Let the Lipschitz constant be as defined.  Then for a coding scheme $f_E$ that
only represents sequences in $S^n = S^n_{[X]\delta} \cup S^n_{[Y]\delta}$, we have that
\[
M \le \frac{\Lip[f_E]}{n} \left( 1 + \delta \diam(\mathcal{G}^n)\right)
\]
\end{thm}
\begin{IEEEproof}
The proof is given as follows:
\begin{align*}
M &= \frac{1}{n} \sum\limits_{x_1^n \in S^n} \sum\limits_{y_1^n \in S^n} p(x_1^n,y_1^n) d_{G^n}(f_E(x_1^n),f_E(y_1^n)) \\ \notag
  &\stackrel{(a)}{\le} \frac{1}{n} \sum\limits_{x_1^n \in S^n} \sum\limits_{y_1^n \in S^n} p(x_1^n,y_1^n) \Lip[f_E]d_{A^n}(x_1^n,y_1^n) \\ \notag
  &= \frac{\Lip[f_E]}{n} \sum\limits_{x_1^n \in S^n} \sum\limits_{y_1^n \in S^n} p(x_1^n,y_1^n) d_{A^n}(x_1^n,y_1^n) \\ \notag
  &= \frac{\Lip[f_E]}{n} \left\{ \sum\limits_{\substack{x_1^n \in S^n, y_1^n \in S^n \\ (x_1^n, y_1^n) \in T^n_{[XY]\delta}}} p(x_1^n,y_1^n) d_{A^n}(x_1^n,y_1^n) + \sum\limits_{\substack{x_1^n \in S^n, y_1^n \in S^n \\ (x_1^n, y_1^n) \notin T^n_{[XY]\delta}}} p(x_1^n,y_1^n) d_{A^n}(x_1^n,y_1^n) \right\} \\ \notag
  &\stackrel{(b)}{=} \frac{\Lip[f_E]}{n} \left\{ \Pr\left[(x_1^n, y_1^n) \in T^n_{[XY]\delta}\right] + \sum\limits_{\substack{x_1^n \in S^n, y_1^n \in S^n \\ (x_1^n, y_1^n) \notin T^n_{[XY]\delta}}} p(x_1^n,y_1^n) d_{A^n}(x_1^n,y_1^n) \right\} \\ \notag
  &\stackrel{(c)}{\le} \frac{\Lip[f_E]}{n} \left\{ 1 + \delta \max\limits_{\substack{x_1^n \in S^n, y_1^n \in S^n \\ (x_1^n, y_1^n) \notin T^n_{[XY]\delta}}} d_{A^n}(x_1^n,y_1^n) \right\} \\ \notag
&= \frac{\Lip[f_E]}{n} \left\{ 1 + \delta \diam(\mathcal{G}^n)\right\} \mbox{,}
\end{align*}
where
step (a) is by definition of the Lipschitz constant; step (b) follows from the definition of graph distance 
and the consistency of strong typicality (\cite[Theorem 5.7]{Yeung2002}); and step (c) is from bounding $\Pr\left[(x_1^n, y_1^n) \in T^n_{[XY]\delta}\right]$ by $1$ and from Lemma~\ref{lem:jaep}.  Note that the $\delta$ bound used in step (c) for the probability of sequences pairs
that are both marginally typical but not jointly typical is actually the probability of all non-jointly typical pairs and is therefore loose.
\end{IEEEproof}

Computing Lipschitz constants is usually difficult or impossible.
There are, however, methods from theoretical computer science for bounding
Lipschitz constants (or dilation) for embeddings \cite{RosenbergH2001,LivingstonS1988}.  For a ``host'' graph
$\mathcal{H}$ and a ``guest'' graph $\mathcal{G}$, a basic counting argument reminiscent of 
Theorem~\ref{thm:expBlockPalimpsest}, shows that the dilation for any embedding of $\mathcal{G}$ into $\mathcal{H}$
must satisfy
\[
\Lip[f_E] \ge \left\lceil \frac{\log(d_{\mathcal{G}} - 1)}{\log(d_{\mathcal{H}} - 1)}\right\rceil \mbox{,}
\]
where $d_{\mathcal{G}}$ and $d_{\mathcal{H}}$ are the respective maximum degrees \cite[Prop. 1.5.2]{RosenbergH2001}.  When the guest graph
is the joint typicality graph $\mathcal{G}^n$ and the host graph is the hypercube $H_{nK}$, this implies
that
\[
\Lip[f_E] \ge \left\lceil \frac{\log(2^{n(H(Y|X) + \nu)} - 1)}{\log(nK - 1)}\right\rceil \approx \frac{nH(Y|X)}{\log nK} \mbox{.}
\]
Another typical result arises when it is fixed that both graphs have $m$ vertices (expansion is $1$). 
The Lipschitz constant bound is in terms the bisection width $W_{\mathcal{G}}$ and the recursive edge-bisector
function $R_{\mathcal{H}}(\cdot)$.  The dilation $\Lip[f_E]$ of any embedding $f_E$ of $\mathcal{G}$ into 
$\mathcal{H}$ must satisfy
\[
\Lip[f_E] \ge \left(\frac{1}{\log d_{\mathcal{H}}}\right)\log \frac{W_{\mathcal{G}}}{d_{\mathcal{G}}R_{\mathcal{H}}(m)} - 1 \mbox{.}
\]
The bisection width of a graph is the size of the smallest cut-set that breaks the graph into two subgraphs of equal 
sizes (to within rounding) \cite[Prop. 2.3.6]{RosenbergH2001}.  Using such a result is difficult since the bisection
width of the joint typicality graph is not known.  For the case when $\mathcal{H} = H_{\lceil nH(X) \rceil}$,
a simplified version reduces to
\[
\Lip[f_E] \ge \frac{\lceil \log m \rceil}{\diam(\mathcal{G}^n)} \approx \frac{nH(X)}{\diam(\mathcal{G}^n)} \mbox{,}
\]
where $\diam(\mathcal{G}^n)$ is the same graph diameter we had seen before \cite{LivingstonS1988}.  If the dilation
is to be no greater than $2$, the PPM scheme we have described previously may be reinterpreted in a graph embedding
framework and seen to achieve $\Lip[f_E] \le 2$, but the price is exponential expansion, $\expan[f_E] = \frac{1}{m}{2^{m-1}}$. 

Returning to Theorem~\ref{thm:palimpsestLipsz}, as noted in Lemma \ref{lem:jaepval}, $\delta$ can be taken as 
\[
\delta(n) = n^{-\frac{1}{2} + \omega}
\]
for some fixed $\omega > 0$. 
The diameter of the hypercube $H_{nK}$ is clearly $nK$.  Combining this 
with the contraction provides a bound on the diameter of $\mathcal{G}^n$:
\[
\diam(\mathcal{G}^n) \le \Lip[f_E^{-1}]nK \mbox{.}
\]
Thus one can further bound the expression in Theorem~\ref{thm:palimpsestLipsz} as
\begin{align*}
M &\le \frac{\Lip[f_E]}{n} \left(1 + \delta \diam(\mathcal{G}^n) \right) \\ \notag
&\le \frac{\Lip[f_E]}{n} \left(1 + n^{-\frac{1}{2} + \omega} \Lip[f_E^{-1}]nK \right) \\ \notag
&= \frac{\Lip[f_E]}{n} + \frac{K \Lip[f_E]\Lip[f_E^{-1}]}{n^{1/2 -\omega}} \mbox{.}
\end{align*}
This yields the following proposition.
\begin{prop}
\label{thm:palimpsestLipsz2}
The malleability $M$ is asymptotically bounded above by:
\[
\limsup_{n \to \infty} M \leq \frac{\Lip[f_E]}{n} + \frac{K\Lip[f_E]\Lip[f_E^{-1}]}{n^{1/2-\omega}}
\]
for any fixed $\omega > 0$.
\end{prop}

The quantity $M$ is essentially bounded by $n^{-1/2}K\Lip[f_E]\Lip[f_E^{-1}]$ since the second term 
should dominate the first.  An alternate expression for $K$ is
$K = \tfrac{1}{n}\log_2 \expan[f_E] + H(X)$, which is fixed.  
If $\Lip[f_E]\Lip[f_E^{-1}]$ is $o(\sqrt{n})$ and $\Lip[f_E]$ is $o(n)$ for the sequence of encoders $f_E$, 
$M$ will go to zero asymptotically in $n$.  Due to the bounding methods that were used,
it is not at all clear whether this Lipschitz bound on malleability is tight, 
and one might suspect that it is not.  A slightly different branch of theoretical computer 
science deals with bounding the distortion of mappings \cite{RabinovichR1998,LinialLR1995}, 
however it is not clear how to apply these
results to the palimpsest problem.

\section{Discussion and Conclusions}
\label{sec:final}
We have formulated an information theoretic problem
motivated by applications in information storage where a compressed stored document
must often be updated and there are costs associated with writing on the
storage medium.
That there are always editing costs in overwriting rewritable media
is a fundamental fact of thermodynamics and follows from Landauer's principle
\cite{BennettGLVZ1998}:
Since discarding information results in a dissipation of energy, overwriting 
causes an inextricable loss of energy.

Both the compressed palimpsest problem considered here and a
distinct problem with a similar motivation presented in a
companion paper~\cite{KusumaVG2008a} exhibit a fundamental trade-off
between compression efficiency and the costs incurred
when synchronizing between two versions of a source code. 
The palimpsest problem is concerned with random access editing,
where changing nearby or greatly separated symbols in the compressed
representation have the same cost.
The ``cut-and-paste'' formulation of~\cite{KusumaVG2008a} is concerned
with editing large subsequences, as would be appropriate when there is
a cost associated with communicating the positions of edits.

The basic result is that unless the two versions of the source
are either very strongly correlated or have a deterministically common part, 
if rates close to entropy are required for both sources, then
a large malleability cost will have to be paid.  Similarly, 
if small malleability is required, a very large rate penalty
will be paid.  There is a fundamental trade-off between the quantities.

For our compressed palimpsest problem,
we found that if minimal malleability costs are desired,
then a rate penalty that is exponential in the conditional entropy
of the editing process must be paid. 
That is, unless the two versions of the source are very strongly correlated
(conditional entropy logarithmic in block length),
rate exponentially larger than entropy is needed. 
A universal scheme for minimal malleability
is given by a pulse position modulation method. 
Thus, if we require malleability $M = O(1/n)$,
then rates $K$ and $L$ must be $\Omega(\tfrac{1}{n}2^n)$.  

One may be tempted to try to cast the block palimpsest problem in terms 
of error-correcting codes, where the quality metric is the block Hamming distance.  
The Hamming distance does not care how two letters differ, it only cares 
whether they are different.  In a sense, it is an $\ell_\infty$ distance.  This gives 
rise to error-correcting codes that try to maximize the minimum distance between 
two codewords in the codebook.  In malleable coding, we care not just about 
whether a modified codeword is inside or outside the minimum distance
decoding region for the original codeword, but how far, basically treating the space
with a symbol edit distance, which may be $\ell_1$.

\appendices
\section{$(\mathcal{V}^{*},d)$ is a Finite Metric Space}
\label{app:prop-edit}
A metric must satisfy non-negativity, equality, symmetry, and the triangle inequality.
These properties are verified for any edit distance with edit operation $R$ as follows.
\begin{itemize}
\item
\emph{non-negativity}: follows since the edit distance is a counting measure.
\item
\emph{equality}: follows by definition, since the distance is zero if and only if $a = b$.
\item
\emph{symmetry}: If $d(a,b) = n$, then it follows there is a sequence of $n - 1$ intermediate 
strings, $a_1, a_2, \ldots, a_{n-1}$ which along with $a_0 = a$ and $a_n = b$ satisfy $(a_i, a_{i+1}) \in R$.  
Since $R$ is a symmetric relation, it follows that $(a_i+1, a_i)$ is also in $R$, and so there 
is a backwards sequence $a_n, a_{n-1}, \ldots, a_0$. Hence if $d(a,b) = n$ then $d(b,a) = n$ 
also, and so $d(a,b) = d(b,a)$ for all $a, b$.
\item
\emph{triangle inequality}: Suppose $d(a, b) + d(b, c) < d(a, c)$. Then there is a 
sequence of editing operations $(a_i, a_{i+1})$ that goes from $a$ to $c$ via $b$ in 
$d(a, b) + d(b, c)$ steps.  Now perform the editing operations of $d(a,b)$ followed 
by the operations of $d(b, c)$, which requires $d(a, b) + d(b, c)$ steps. 
This contradicts the initial assumption, hence $d(a, b) + d(b, c) \ge d(a, c)$.
\end{itemize}

\section{Proof of Proposition~\ref{prop:cartesian-product}}
\label{app:prop-cartesian-product}
Since $G_1 \leadsto H_1$, $V(G_1) \subseteq V(H_1)$.  
Since $G_2 \leadsto H_2$, $V(G_2) \subseteq V(H_2)$.
Then by elementary set operations,
$V(G_1\times G_2) = V(G_1) \times V(G_2) \subseteq V(H_1\times H_2) = V(H_1) \times V(H_2)$.  
Since $G_1 \leadsto H_1$, $E(G_1) \subseteq E(H_1)$.  
Since $G_2 \leadsto H_2$, $E(G_2) \subseteq E(H_2)$.  
Consider an edge 
$(u,v) \in E(G_1\times G_2)$.  By definition of Cartesian product, it satisfies 
($u_1=v_1$ and $(u_2,v_2) \in E(G_2)$) or ($u_2=v_2$ and $(u_1,v_1) \in E(G_1)$),
but since $E(G_1) \subseteq E(H_1)$ and $E(G_2) \subseteq E(H_2)$, 
it also satisfies ($u_1=v_1$ and $(u_2,v_2) \in E(H_2)$) or ($u_2=v_2$ and $(u_1,v_1) \in E(H_1)$).
Therefore $E(G_1\times G_2) \subseteq E(H_1\times H_2)$. 
Since $V(G_1\times G_2) \subseteq V(H_1\times H_2)$ and $E(G_1\times G_2) \subseteq E(H_1\times H_2)$,
$G_1\times G_2 \leadsto H_1\times H_2$.

\section*{Acknowledgments}
The second author thanks Vahid Tarokh for introducing him to storage
area networks.  The authors also thank Robert G. Gallager and Sanjoy 
K. Mitter for useful exchanges; Sekhar Tatikonda for discussions
on mappings between source and representation spaces; and
Renuka K. Sastry for assistance with genetics.

\bibliographystyle{IEEEtran} 
\bibliography{abrv,lrv_lib}

\begin{thebibliography}{10}
\providecommand{\url}[1]{#1}
\csname url@rmstyle\endcsname
\providecommand{\newblock}{\relax}
\providecommand{\bibinfo}[2]{#2}
\providecommand\BIBentrySTDinterwordspacing{\spaceskip=0pt\relax}
\providecommand\BIBentryALTinterwordstretchfactor{4}
\providecommand\BIBentryALTinterwordspacing{\spaceskip=\fontdimen2\font plus
\BIBentryALTinterwordstretchfactor\fontdimen3\font minus
  \fontdimen4\font\relax}
\providecommand\BIBforeignlanguage[2]{{%
\expandafter\ifx\csname l@#1\endcsname\relax
\typeout{** WARNING: IEEEtran.bst: No hyphenation pattern has been}%
\typeout{** loaded for the language `#1'. Using the pattern for}%
\typeout{** the default language instead.}%
\else
\language=\csname l@#1\endcsname
\fi
#2}}

\bibitem{Shannon1948}
C.~E. Shannon, ``A mathematical theory of communication,'' \emph{Bell Syst.
  Tech. J.}, vol.~27, pp. 379--423, 623--656, July/Oct. 1948.

\bibitem{Huffman1952}
D.~A. Huffman, ``A method for the construction of minimum-redundancy codes,''
  \emph{Proc. {IRE}}, vol.~40, no.~9, pp. 1098--1101, Sept. 1952.

\bibitem{NetzN2007}
R.~Netz and W.~Noel, \emph{The {A}rchimedes Codex}.\hskip 1em plus 0.5em minus
  0.4em\relax Philadelphia, PA: Da Capo Press, 2007.

\bibitem{KusumaVG2008a}
J.~Kusuma, L.~R. Varshney, and V.~K. Goyal, ``Malleable coding: A cut-and-paste
  method,'' \emph{{IEEE} Trans. Inf. Theory}, 2008, in preparation.

\bibitem{MessmerB1998}
B.~T. Messmer and H.~Bunke, ``A new algorithm for error-tolerant subgraph
  isomorphism detection,'' \emph{{IEEE} Trans. Pattern Anal. Mach. Intell.},
  vol.~20, no.~5, pp. 493--504, May 1998.

\bibitem{Shannon1956}
C.~E. Shannon, ``The zero error capacity of a noisy channel,'' \emph{{IRE}
  Trans. Inf. Theory}, vol. IT-2, no.~3, pp. 8--19, Sept. 1956.

\bibitem{Witsenhausen1976}
H.~S. Witsenhausen, ``The zero-error side information problem and chromatic
  numbers,'' \emph{{IEEE} Trans. Inf. Theory}, vol. IT-22, no.~5, pp. 592--593,
  Sept. 1976.

\bibitem{GrossT1987}
J.~L. Gross, \emph{Topological Graph Theory}.\hskip 1em plus 0.5em minus
  0.4em\relax New York: John Wiley \& Sons, 1987.

\bibitem{Tlusty2007}
T.~Tlusty, ``A model for the emergence of the genetic code as a transition in a
  noisy information channel,'' \emph{J. Theor. Biol.}, vol. 249, no.~2, pp.
  331--342, Nov. 2007.

\bibitem{BobbarjungJD2006}
D.~R. Bobbarjung, S.~Jagannathan, and C.~Dubnicki, ``Improving duplicate
  elimination in storage systems,'' \emph{ACM Trans. Storage}, vol.~2, no.~4,
  pp. 424--448, Nov. 2006.

\bibitem{PolicroniadesP2004}
C.~Policroniades and I.~Pratt, ``Alternatives for detecting redundancy in
  storage systems data,'' in \emph{Proc. 2004 {USENIX} Ann. Tech. Conf.},
  Boston, June 2004, pp. 73--86.

\bibitem{BurnsSL2003}
R.~Burns, L.~Stockmeyer, and D.~D.~E. Long, ``In-place reconstruction of
  version differences,'' \emph{{IEEE} Trans. Knowl. Data Eng.}, vol.~15, no.~4,
  pp. 973--984, July-Aug. 2003.

\bibitem{SuelM2003}
T.~Suel and N.~Memon, ``Algorithms for delta compression and remote file
  synchronization,'' in \emph{Lossless Compression Handbook}, K.~Sayood,
  Ed.\hskip 1em plus 0.5em minus 0.4em\relax Elsevier, 2003, pp. 269--290.

\bibitem{MitterN2005}
S.~K. Mitter and N.~J. Newton, ``Information and entropy flow in the
  {K}alman-{B}ucy filter,'' \emph{J. Stat. Phys.}, vol. 118, no. 1-2, pp.
  145--176, Jan. 2005.

\bibitem{AhlswedeZ1989}
R.~Ahlswede and Z.~Zhang, ``Coding for write-efficient memory,'' \emph{Inf.
  Comput.}, vol.~83, no.~1, pp. 80--97, Oct. 1989.

\bibitem{AhlswedeZ1994}
------, ``On multiuser write-efficient memories,'' \emph{{IEEE} Trans. Inf.
  Theory}, vol.~40, no.~3, pp. 674--686, May 1994.

\bibitem{RamprasadSH1999c}
S.~Ramprasad, N.~R. Shanbhag, and I.~N. Hajj, ``Information-theoretic bounds on
  average signal transition activity,'' \emph{{IEEE} Trans. {VLSI} Syst.},
  vol.~7, no.~3, pp. 359--368, Sept. 1999.

\bibitem{Orlitsky1993}
A.~Orlitsky, ``Interactive communication of balanced distributions and of
  correlated files,'' \emph{SIAM J. Discrete Math.}, vol.~6, no.~4, pp.
  548--564, Nov. 1993.

\bibitem{MinskyTZ2003}
Y.~Minsky, A.~Trachtenberg, and R.~Zippel, ``Set reconciliation with nearly
  optimal communication complexity,'' \emph{{IEEE} Trans. Inf. Theory},
  vol.~49, no.~9, pp. 2213--2218, Sept. 2003.

\bibitem{GarfinkelEEF2007}
\BIBentryALTinterwordspacing
M.~S. Garfinkel, D.~Endy, G.~L. Epstein, and R.~M. Friedman, ``Synthetic
  genomics: Options for governance,'' Oct. 2007. [Online]. Available:
  \url{http://hdl.handle.net/1721.1/39141}
\BIBentrySTDinterwordspacing

\bibitem{WongWF2003}
P.~C. Wong, K.-K. Wong, and H.~Foote, ``Organic data memory using the {DNA}
  approach,'' \emph{Commun. ACM}, vol.~46, no.~1, pp. 95--98, Jan. 2003.

\bibitem{Swanson1984}
R.~Swanson, ``A unifying concept for the amino acid code,'' \emph{Bull. Math.
  Biol.}, vol.~46, no.~2, pp. 187--203, Mar. 1984.

\bibitem{PrimroseTO2001}
S.~B. Primrose, R.~M. Twyman, and R.~W. Old, \emph{Principles of Gene
  Manipulation}, 6th~ed.\hskip 1em plus 0.5em minus 0.4em\relax Oxford:
  Blackwell Science, 2001.

\bibitem{VarshneySC2006}
L.~R. Varshney, P.~J. {Sj\"{o}str\"{o}m}, and D.~B. Chklovskii, ``Optimal
  information storage in noisy synapses under resource constraints,''
  \emph{Neuron}, vol.~52, no.~3, pp. 409--423, Nov. 2006.

\bibitem{Cormode2003}
G.~Cormode, ``Sequence distance embeddings,'' Ph.D. dissertation, University of
  Warwick, Jan. 2003.

\bibitem{Gilbert1971}
E.~N. Gilbert, ``Codes based on inaccurate source probabilities,'' \emph{{IEEE}
  Trans. Inf. Theory}, vol. IT-17, no.~3, pp. 304--314, May 1971.

\bibitem{Davisson1973}
L.~D. Davisson, ``Universal noiseless coding,'' \emph{{IEEE} Trans. Inf.
  Theory}, vol. IT-19, no.~6, pp. 783--795, Nov. 1973.

\bibitem{Topsoe2000}
F.~Tops{\o}e, ``Some inequalities for information divergence and related
  measures of discrimination,'' \emph{{IEEE} Trans. Inf. Theory}, vol.~46,
  no.~4, pp. 1602--1609, July 2000.

\bibitem{SinanovicJ2007}
S.~{Sinanovi\'{c}} and D.~H. Johnson, ``Toward a theory of information
  processing,'' \emph{Signal Process.}, vol.~87, no.~6, pp. 1326--1344, June
  2007.

\bibitem{DavissonMPW1981}
L.~D. Davisson, R.~J. McEliece, M.~B. Pursley, and M.~S. Wallace, ``Efficient
  universal noiseless source codes,'' \emph{{IEEE} Trans. Inf. Theory}, vol.
  IT-27, no.~3, pp. 269--279, May 1981.

\bibitem{GacsK1973}
P.~G{\'{a}}cs and J.~K{\"{o}}rner, ``Common information is far less than mutual
  information,'' \emph{Probl. Control Inf. Theory}, vol.~2, no.~2, pp.
  149--162, 1973.

\bibitem{VisweswariahKV1998}
K.~Visweswariah, S.~R. Kulkarni, and S.~{Verd\'{u}}, ``Source codes as random
  number generators,'' \emph{{IEEE} Trans. Inf. Theory}, vol.~44, no.~2, pp.
  462--471, Mar. 1998.

\bibitem{FuS1997}
F.~Fu and S.~Shen, ``On the expectation and variance of {H}amming distance
  between two i.i.d. random vectors,'' \emph{Acta Math. Appl. Sin.}, vol.~13,
  no.~3, pp. 243--250, July 1997.

\bibitem{Korner1971}
J.~{K\"{o}rner}, ``A property of conditional entropy,'' \emph{Stud. Sci. Math.
  Hung.}, vol.~6, pp. 355--359, 1971.

\bibitem{EquitzC1991}
W.~H.~R. Equitz and T.~M. Cover, ``Successive refinement of information,''
  \emph{{IEEE} Trans. Inf. Theory}, vol.~37, no.~2, pp. 269--275, Mar. 1991.

\bibitem{Verdu1990}
S.~{Verd\'{u}}, ``On channel capacity per unit cost,'' \emph{{IEEE} Trans. Inf.
  Theory}, vol.~36, no.~5, pp. 1019--1030, Sept. 1990.

\bibitem{VarshneyG2006b}
L.~R. Varshney and V.~K. Goyal, ``Ordered and disordered source coding,'' in
  \emph{Proc. Inf. Theory Appl. Inaugural Workshop}, La Jolla, California, Feb.
  2006.

\bibitem{PradhanCR2006}
S.~S. Pradhan, S.~Choi, and K.~Ramchandran, ``A graph-based framework for
  transmission of correlated sources over multiple-access channels,''
  \emph{{IEEE} Trans. Inf. Theory}, vol.~53, no.~12, pp. 4583--4604, Dec. 2007.

\bibitem{AgrellLSO2004}
E.~Agrell, J.~Lassing, E.~G. {Str\"{o}m}, and T.~Ottosson, ``On the optimality
  of the binary reflected {G}ray code,'' \emph{{IEEE} Trans. Inf. Theory},
  vol.~50, no.~12, pp. 3170--3182, Dec. 2004.

\bibitem{LongoG1982}
G.~Longo and G.~Galasso, ``An application of informational divergence to
  {H}uffman codes,'' \emph{{IEEE} Trans. Inf. Theory}, vol. IT-28, no.~1, pp.
  36--43, Jan. 1982.

\bibitem{Gilbert1958}
E.~N. Gilbert, ``Gray codes and paths on the $n$-cube,'' \emph{Bell Syst. Tech.
  J.}, vol.~37, no.~3, pp. 815--826, May 1958.

\bibitem{Dobrushin1967}
R.~L. Dobrushin, ``{Shannon's} theorems for channels with synchronization
  errors,'' \emph{Probl. Inf. Transm.}, vol.~3, no.~4, pp. 11--26, Oct.-Dec.
  1967.

\bibitem{ZegerG1990}
K.~Zeger and A.~Gersho, ``Pseudo-{G}ray coding,'' \emph{{IEEE} Trans. Commun.},
  vol.~38, no.~12, pp. 2147--2158, Dec. 1990.

\bibitem{CaireTB1998}
G.~Caire, G.~Taricco, and E.~Biglieri, ``Bit-interleaved coded modulation,''
  \emph{{IEEE} Trans. Inf. Theory}, vol.~44, no.~3, pp. 927--946, May 1998.

\bibitem{SethuramanH2006}
V.~Sethuraman and B.~Hajek, ``Comments on ``bit-interleaved coded
  modulation'','' \emph{{IEEE} Trans. Inf. Theory}, vol.~52, no.~4, pp.
  1795--1797, Apr. 2006.

\bibitem{Losee1992}
R.~M. Losee, Jr., ``A {G}ray code based ordering for documents on shelves:
  Classification for browsing and retrieval,'' \emph{J. Am. Soc. Inform. Sci.},
  vol.~43, no.~4, pp. 312--322, May 1992.

\bibitem{Edelman1999}
S.~Edelman, \emph{Representation and Recognition in Vision}.\hskip 1em plus
  0.5em minus 0.4em\relax Cambridge: MIT Press, 1999.

\bibitem{AlonO1996}
N.~Alon and A.~Orlitsky, ``Source coding and graph entropies,'' \emph{{IEEE}
  Trans. Inf. Theory}, vol.~42, no.~5, pp. 1329--1339, Sept. 1996.

\bibitem{KornerO1998}
J.~{K\"{o}rner} and A.~Orlitsky, ``Zero-error information theory,''
  \emph{{IEEE} Trans. Inf. Theory}, vol.~44, no.~6, pp. 2207--2229, Oct. 1998.

\bibitem{GareyJ1979}
M.~R. Garey and D.~S. Johnson, \emph{Computers and Intractability: A Guide to
  the Theory of {NP}-Completeness}.\hskip 1em plus 0.5em minus 0.4em\relax San
  Francisco: W.~H. Freeman, 1979.

\bibitem{KoschutzkiLPRTZ2005}
D.~{Kosch\"{u}tzki}, K.~A. Lehmann, L.~Peeters, S.~Richter, D.~Tenfelde-Podehl,
  and O.~Zlotowski, ``Centrality indices,'' in \emph{Network Analysis:
  Methodological Foundations}, U.~Brandes and T.~Erlebach, Eds.\hskip 1em plus
  0.5em minus 0.4em\relax Berlin: Springer, 2005, pp. 16--61.

\bibitem{Wiener1947}
H.~Wiener, ``Structural determination of paraffin boiling points,'' \emph{J.
  Am. Chem. Soc.}, vol.~69, no.~1, pp. 17--20, Jan. 1947.

\bibitem{AusielloISN1991}
G.~Ausiello, G.~F. Italiano, A.~M. Spaccamela, and U.~Nanni, ``Incremental
  algorithms for minimal length paths,'' \emph{J. Algorithms}, vol.~12, no.~4,
  pp. 615--638, Dec. 1991.

\bibitem{DemetrescuI2004}
C.~Demetrescu and G.~F. Italiano, ``A new approach to dynamic all pairs
  shortest paths,'' \emph{J. ACM}, vol.~51, no.~6, pp. 968--992, Nov. 2004.

\bibitem{Bunke1999}
H.~Bunke, ``Error correcting graph matching: On the influence of the underlying
  cost function,'' \emph{{IEEE} Trans. Pattern Anal. Mach. Intell.}, vol.~21,
  no.~9, pp. 917--922, Sept. 1999.

\bibitem{Kraft1949}
L.~G. Kraft, Jr., ``A device for quantizing, grouping, and coding
  amplitude-modulated pulses,'' Master's thesis, Massachusetts Institute of
  Technology, 1949.

\bibitem{Ahlswede2006}
R.~Ahlswede, ``Identification entropy,'' in \emph{General Theory of Information
  Transfer and Combinatorics}, ser. Lecture Notes in Computer Science,
  R.~Ahlswede, L.~{B\"{a}umer}, N.~Cai, H.~Aydinian, V.~Blinovsky, C.~Deppe,
  and H.~Mashurian, Eds.\hskip 1em plus 0.5em minus 0.4em\relax Berlin:
  Springer, 2006, vol. 4123, pp. 595--613.

\bibitem{Yeung2002}
R.~W. Yeung, \emph{A First Course in Information Theory}.\hskip 1em plus 0.5em
  minus 0.4em\relax New York: Kluwer Academic/Plenum Publishers, 2002.

\bibitem{CsiszarK1997}
I.~{Csisz\'{a}r} and J.~{K\"{o}rner}, \emph{Information Theory: Coding Theorems
  for Discrete Memoryless Systems}, 3rd~ed.\hskip 1em plus 0.5em minus
  0.4em\relax Budapest: {Akad\'{e}miai} {Kiad\'{o}}, 1997.

\bibitem{DezaL1997}
M.~M. Deza and M.~Laurent, \emph{Geometry of Cuts and Metrics}.\hskip 1em plus
  0.5em minus 0.4em\relax Berlin: Springer, 1997.

\bibitem{DezaGS2004}
M.~Deza, V.~Grishukhin, and M.~Shtogrin, \emph{Scale-Isometric Polytopal Graphs
  in Hypercubes and Cubic Lattices}.\hskip 1em plus 0.5em minus 0.4em\relax
  London: Imperial College Press, 2004.

\bibitem{RosenbergH2001}
A.~L. Rosenberg and L.~S. Heath, \emph{Graph Separators, with
  Applications}.\hskip 1em plus 0.5em minus 0.4em\relax New York: Kluwer
  Academic / Plenum Publishers, 2001.

\bibitem{LivingstonS1988}
M.~Livingston and Q.~F. Stout, ``Embeddings in hypercubes,'' \emph{Math.
  Comput. Model.}, vol.~11, pp. 222--227, 1988.

\bibitem{RabinovichR1998}
Y.~Rabinovich and R.~Raz, ``Lower bounds on the distortion of embedding finite
  metric spaces in graphs,'' \emph{Discrete Comput. Geom.}, vol.~19, no.~1, pp.
  79--94, Jan. 1998.

\bibitem{LinialLR1995}
N.~Linial, E.~London, and Y.~Rabinovich, ``The geometry of graphs and some of
  its algorithmic applications,'' \emph{Combinatorica}, vol.~15, no.~2, pp.
  215--245, June 1995.

\bibitem{BennettGLVZ1998}
C.~H. Bennett, P.~{G\'{a}cs}, M.~Li, P.~M.~B. {Vit\'{a}nyi}, and W.~H. Zurek,
  ``Information distance,'' \emph{{IEEE} Trans. Inf. Theory}, vol.~44, no.~4,
  pp. 1407--1423, July 1998.

\end{thebibliography}
\end{document}